\documentclass[final, 3p, times]{elsarticle}

\usepackage{graphicx, hyperref, xspace, float}
\usepackage{amsmath, amsfonts, amssymb, xcolor, comment, bm}
\usepackage[caption=false]{subfig}

\hypersetup{colorlinks,
            linkcolor = {red!50!black},
            citecolor = {red!10!black},
            urlcolor  = {green!10!black}}

\journal{Physical Review E}

\begin{document}

\begin{frontmatter}



\title{Exponential methods for anisotropic diffusion}


\author[inst1]{Pranab J. Deka}\corref{lod1}
\ead{pranab.deka@uibk.ac.at}
\cortext[lod1]{Corresponding author}
\author[inst1]{Lukas Einkemmer}
\affiliation[inst1]{organization = {Department of Mathematics, University of Innsbruck},
            city = {Innsbruck},
            postcode = {A-6020}, 
            country = {Austria}}
            
\author[inst2]{Ralf Kissmann}
\affiliation[inst2]{organization = {Institute for Astro- and Particle Physics, University of Innsbruck},
            city = {Innsbruck},
            postcode = {A-6020}, 
            country = {Austria}}

\begin{abstract}

The anisotropic diffusion equation is imperative in understanding cosmic ray diffusion across the Galaxy, the heliosphere, and its interplay with the ambient magnetic field. This diffusion term contributes to the highly stiff nature of the CR transport equation. In order to conduct numerical simulations of time-dependent cosmic ray transport, implicit integrators have been traditionally favoured over the CFL-bound explicit integrators in order to be able to take large step sizes. We propose exponential methods that directly compute the exponential of the matrix to solve the linear anisotropic diffusion equation. These methods allow us to take even larger step sizes; in certain cases, we are able to choose a step size as large as the simulation time, i.e., only one time step. This can substantially speed-up the simulations whilst generating highly accurate solutions (l2 error $\leq 10^{-10}$). Additionally, we test an approach based on extracting a constant diffusion coefficient from the anisotropic diffusion equation, where the constant coefficient term is solved implicitly or exponentially and the remainder is treated using some explicit method. We find that this approach, for homogeneous linear problems, is unable to improve on the exponential-based methods that directly evaluate the matrix exponential. 
\end{abstract}



\begin{keyword}
Anisotropic Diffusion \sep Cosmic rays \sep Matrix exponential \sep Exponential Integrators \sep Semi-implicit schemes
\end{keyword}

\end{frontmatter}


\newcommand{\pjd}[1]{{{\color{blue} #1}}}

\newcommand{\aap}{A\&A}
\newcommand{\apss}{Astro. \& Astrophys.dics Suppl.}
\newcommand{\aaps}{Astro. \& Astrophys. Suppl.}
\newcommand{\acp}{Atmos. Chem. \& Phys.}
\newcommand{\acpd}{Atmos. Chem. \& Phys. Discus.}
\newcommand{\ag}{Annales Geophysicae}
\newcommand{\aip}{AIP}
\newcommand{\aj}{Astron. J.}
\newcommand{\apj}{ApJ}
\newcommand{\apjl}{ApJL}
\newcommand{\apjs}{ApJS}
\newcommand{\araa}{Annual. Rev. Astron. Astrophys.}
\newcommand{\asr}{Adv. Space Res.}
\newcommand{\aisr}{Adv. Space Res.dearch}
\newcommand{\grl}{Geophys. Res. Lett.}
\newcommand{\ICRC}{Internat. Cos. Ray Conf.}
\newcommand{\jastp}{J. Atmos. Solar-Terr. Phys.}
\newcommand{\jcp}{J. Chem. Phys.}
\newcommand{\jgr}{J. Geophys. Res.}
\newcommand{\mnras}{MNRAS}
\newcommand{\nat}{Nature}
\newcommand{\pasj}{Publ. Astron. Soc. Japan}
\newcommand{\physrep}{Phys. Rep.}
\newcommand{\physscr}{Physica Scripta}
\newcommand{\planss}{Planet. Space Sci.}
\newcommand{\prd}{Phys.~Rev.~D} 
\newcommand{\pre}{Phys.~Rev.~E} 
\newcommand{\prl}{Phys.~Rev.~Lett.} 
\newcommand{\solphys}{Sol. Phys.}
\newcommand{\ssr}{Space Sci. Rev.}
\newcommand{\nar}{New~Astron.~Rev.}
\newcommand{\jcap}{J. Cosmol. Astropart. Phys.}


\newcommand{\picard}{\textsc{Picard}\xspace}


\section{Introduction}

If a certain scalar physical quantity diffuses preferentially along a certain direction, then the diffusion is said to be anisotropic. Anisotropic diffusion is ubiquitous in many fields of science and engineering such as geological systems \cite{Berkowitz02, Saadatfar02, VanLoon04}, material physics \cite{Watson10, Li17}, biology \cite{Pawar14}, image processing, \cite{Perona90, Weickert98}, magnetic resonance imaging \cite{Basser02, Palma14}, and plasma physics \cite{Braginskii65, Holod05}. Anisotropic diffusion also finds applications in several astrophysical scenarios. Anisotropic thermal conduction, arising from electrons (and possibly ions) moving preferentially along the magnetic field lines, is of importance in studying the dynamics and stability of the magnetised intracluster medium \cite{Parrish05, Parrish08, Parrish09}. An appropriate treatment of heat conduction is essential in understanding the dynamical state of the galaxy clusters \cite{Balbus00}. Highly energetic charged particles of astrophysical origin, known as cosmic rays (CRs), are transported from their sources to different parts of the Galaxy, primarily through diffusion \cite{Strong07, Kachelriess19}. This diffusion depends on the structure, orientation, and strength of the Galactic magnetic field. The Galactic magnetic field can loosely be divided into the large-scale ordered magnetic field and the small-scale turbulent fields. The magnitude of the large-scale ordered field is larger than that of the small-scale fields, thereby resulting in preferential diffusion of the CRs along the ordered magnetic field lines, i.e., CR diffusion is anisotropic \cite{Strong07, Kachelriess19}. The significance of anisotropy of CR diffusion has been recognised \cite{Hanasz09, Shalchi09, Effenberger12, Pakmor16a, Cerri2017, Kachelriess19}, yet the interplay physics between (anisotropic) CR diffusion and the Galactic magnetic field has been poorly understood owing to the paucity of CR observational data as well as the lack of our knowledge of the magnetic field of our Galaxy. Furthermore, studies modelling the transport of CRs in the heliosphere found that anisotropic diffusion is also of crucial importance in understanding the distribution of CRs inside the solar system \cite{FerreiraEtAl2001JGR106_29313, Schlickeiser02, Shalchi09, Potgieter2013LRSP10_3}.

The \picard code was developed by \citet{Kissmann14} to numerically solve the CR transport equation \cite{Strong07}. Although it has been optimised to solve the steady-state transport equation, it is well-suited to handle time-dependent problems. The aim of this paper, to develop efficient solvers for the anisotropic diffusion problem, stems from our desire to study and efficiently treat anisotropic CR diffusion in the Galaxy using \picard. 

Conducting numerical simulations of diffusion-dominated physical phenomena presents a number of challenges. One such challenge is the stringent Courant--Friedrich--Lewy (CFL) condition that renders explicit temporal integrators highly ineffective. Owing to stability constraints, they are forced to choose extremely small time step sizes thereby resulting in large computational runtimes. Implicit or semi-implicit methods have been favoured for temporal integration of these problems as they do not suffer from such restrictions based on stability constraints. A directionally-split semi-implicit method has been proposed in Ref. \cite{Sharma11} which is stable for large step sizes. Semi-implicit integrators have been tested on unstructured meshes in Ref. \cite{Pakmor16b}. Fully implicit schemes with higher-order finite volume methods have been analysed in Ref. \cite{Gunter07}. Certain semi-implicit methods allow one to consider the (dominant) diffusion term implicitly and the remainder of the terms explicitly. The effectiveness of such methods has been demonstrated in Refs. \cite{Filbet10, Filbet12, Crouseilles15}. However, implicit or semi-implicit schemes are subject to a certain order of accuracy. Even for linear problems, they always incur a time discretization error which may restrict the time step size that can be used, especially for lower-order schemes. 

An alternative to using implicit or semi-implicit methods for solving the linear anisotropic diffusion equation is to directly compute the exponential of the diffusion matrix. In doing so, one can obtain the \textit{exact} solution in time (subject to the error incurred during spatial discretisation). For small matrices, the computation of the  exponential of a matrix can be trivially accomplished by means of Pad\'e approximation or eigenvalue decomposition. For large matrices, however, these methods are computationally prohibitive. In this work, we investigate the performance of two computationally attractive methods that compute the exponential of a matrix: the method of polynomial interpolation at Leja points \cite{Caliari04} and the $\mu$-mode integrator \cite{Caliari22a}. We note that the Leja interpolation scheme is an iterative method that can be used to compute the exponential of any matrix whilst the applicability of the direct solver, the $\mu$--mode integrator, is restricted to matrices that can be expressed in the form of the Kronecker sum. Our aim is to show that these exponential methods are far superior to any implicit or semi-implicit approaches both in terms of computational cost incurred as well as the accuracy of the solutions. This paper is structured as follows: in Sec. \ref{sec:spatial}, we outline the spatial discretisation used (for the mixed derivatives). We outline the different temporal integrators and the approaches used for numerically treating the anisotropic diffusion equation in Sec. \ref{sec:temporal}. We describe the test problems under consideration in Sec. \ref{sec:test}, present the results in Sec. \ref{sec:results}, and finally conclude in Sec. \ref{sec:conclude}.


\section{Spatial Discretization}
\label{sec:spatial}

Let us consider the two-dimensional anisotropic diffusion equation (in the absence of any external sources)
\begin{equation}
    \frac{\partial u}{\partial t} = \nabla \cdot (\mathcal{D} \, \nabla \, u(x, y, t))
    \label{eq:anidiff}
\end{equation}
where
\begin{equation}
    \mathcal{D} =   \begin{bmatrix}
                        D_{xx}(x, y) \quad     D_{xy}(x, y)	\\
                        D_{yx}(x, y) \quad     D_{yy}(x, y) \\
                    \end{bmatrix}
\end{equation}
describes the diffusion tensor: $D_{xx}(x, y)$ and $D_{yy}(x, y)$ are the diffusion coefficients along X-- and Y--directions, respectively. We will refer to $D_{xy}(x, y)$ and $D_{yx}(x, y)$ as `drift' coefficients that give rise to the mixed derivatives in the diffusion equation. Eq. \eqref{eq:anidiff} can be expanded as
\begin{equation}
    \frac{\partial u}{\partial t} = \frac{\partial}{\partial x} \left(D_{xx} \frac{\partial u}{\partial x}\right) + \frac{\partial}{\partial y} \left(D_{yy} \frac{\partial u}{\partial y}\right) + \frac{\partial}{\partial x} \left(D_{xy} \frac{\partial u}{\partial y}\right) + \frac{\partial}{\partial y} \left(D_{yx} \frac{\partial u}{\partial x}\right).
    \label{eq:anidiff_discrete}
\end{equation}
It is to be noted that diffusion can also be off-diagonal if the coordinate system is not aligned with the magnetic field lines. We consider sufficiently smooth solutions and adopt the assumption $D_{xy}(x, y) = D_{yx}(x, y)$ which also allows one to interchange the order of differentiation in the mixed-derivative terms. After discretisation using second-order finite differences, the above equation can be written in Kronecker form as
\begin{equation}
    \frac{\partial u}{\partial t} = (I_y \otimes A_{xx}) + (A_{yy} \otimes I_x) + (A_x \otimes A_y) + (A_y \otimes A_x),
\end{equation}
where $A_{xx}$ and $A_{yy}$ are the Laplacian matrices corresponding to $\frac{\partial}{\partial x} \left(D_{xx} \frac{\partial u}{\partial x}\right)$ and $\frac{\partial}{\partial y} \left(D_{y} \frac{\partial u}{\partial y}\right)$, respectively. $A_x$ and $A_y$ are matrices arising from the discretisation of the last two terms on the right-hand-side (RHS) of Eq. \eqref{eq:anidiff_discrete}, respectively. Whilst the discretisation of the first two terms is trivial, the discretisation equations for the mixed-derivative terms are as follows: 
\begin{align*}
    & \frac{\partial}{\partial x} \left(D_{xy} \frac{\partial u}{\partial y}\right) + \frac{\partial}{\partial y} \left(D_{xy} \frac{\partial u}{\partial x}\right) \nonumber \\
    = & \frac{\partial}{\partial x} \left(D_{xy} \frac{u_{i, j+1} - u_{i, j-1}}{2 \Delta y}\right) + \frac{\partial}{\partial y} \left(D_{xy} \frac{u_{i+1, j} - u_{i-1, j}}{2 \Delta x}\right) \nonumber \\
    = & \frac{1}{2 \Delta y} \left[D_{xy, i, j+1} \left(\frac{u_{i+1, j+1} - u_{i-1, j+1}}{2 \Delta x}\right) - D_{xy, i, j-1}\left(\frac{u_{i+1, j-1} - u_{i-1, j-1}}{2 \Delta x}\right) \right] \nonumber \\
    + & \frac{1}{2 \Delta x} \left[D_{xy, i+1, j} \left(\frac{u_{i+1, j+1} - u_{i+1, j-1}}{2 \Delta y}\right) - D_{xy, i-1, j}\left(\frac{u_{i-1, j+1} - u_{i-1, j-1}}{2 \Delta y}\right) \right] \nonumber \\
    = & \frac{1}{4 \Delta x \Delta y} \left[D_{xy, i, j+1} u_{i+1, j+1} - D_{xy, i, j+1} u_{i-1, j+1} - D_{xy, i, j-1} u_{i+1, j-1} + D_{xy, i, j-1} u_{i-1, j-1} \right] \nonumber \\
    + & \frac{1}{4 \Delta x \Delta y} \left[D_{xy, i+1, j} u_{i+1, j+1} - D_{xy, i+1, j} u_{i+1, j-1} - D_{xy, i-1, j} u_{i-1, j+1} + D_{xy, i-1, j} u_{i-1, j-1} \right]
\end{align*}


\section{Temporal Integration}
\label{sec:temporal}

In this section, we describe the different temporal integrators and the approaches used in this study.


\subsection{Implicit integrators}

The well-known second-order implicit integrator, Crank--Nicolson,
\begin{equation*}
    u^{n+1} = u^n + \frac{\Delta t}{2} \left(\nabla \cdot (\mathcal{D} \nabla u^n) + \nabla \cdot (\mathcal{D} \nabla u^{n+1}) \right),
\end{equation*}
has been typically favoured for solving the time-dependent CR transport equation. We use the GMRES module available in \texttt{scipy.sparse.linalg}, in \textsc{Python}, as the iterative solver. We did not obtain any significant performance difference between GMRES and conjugate gradient, and since GMRES is not limited to symmetric matrices, we favour this iterative scheme over others in this work. The performance of Crank--Nicolson is compared and contrasted with our proposed method of directly computing the matrix exponential using the Leja polynomial method. 


\subsection{Semi-implicit integrators}

We adopt the approach of penalisation strategy proposed in Refs. \cite{Filbet10, Filbet12} and was subsequently used in the context of the anisotropic diffusion problem in Ref. \cite{Crouseilles15}. We penalise the anisotropic diffusion equation by the Laplacian operator with a constant diffusion coefficient ($\lambda$). Following Ref. \cite{Crouseilles15}, we choose $\lambda > \alpha$, where $\alpha$ is the largest eigenvalue of $\mathcal{D}$:
\begin{align*}
    \frac{\partial u}{\partial t} & = \lambda \nabla^2 u + \left(\nabla \cdot (\mathcal{D} \nabla u) - \lambda \nabla^2 u \right) \nonumber \\
    \implies \frac{u^{n+1} - u^u}{\Delta t} & = \lambda \nabla^2 u^{n+1} + \left(\nabla \cdot (\mathcal{D} \nabla u^n) - \lambda \nabla^2 u^n \right) \nonumber \\
    \implies (I - \lambda \nabla^2 \Delta t)\,u^{n+1} & = u^n + \left(\nabla \cdot (\mathcal{D} \nabla u^n) - \lambda \nabla^2 u^n \right) \Delta t
\end{align*}
Using implicit-explicit methods, one can solve the stiff term ($\lambda \nabla^2$) implicitly whilst treating the non-stiff remainder $\left(\nabla \cdot (\mathcal{D} \nabla u) - \lambda \nabla^2 u \right)$ explicitly. In this work, we consider a specific class of the implicit-explicit methods -- the additive Runge-Kutta (ARK) schemes \citep{Araujo97, Kennedy03, Higueras06, Liu06, Zhang12}, the gist of which follows. Let us consider an equation of the form
\begin{equation}
	\frac{du}{dt} = f(u(t)) + g(u(t)),
	\label{eq:imex}
\end{equation}
where $f(u(t))$ is the linear and/or stiff term and $g(u(t))$ is the nonlinear and/or non-stiff term. The $i^\mathrm{th}$ stage of an ARK integrator reads
\begin{align*}
	U_i & = u^n + \Delta t \sum_{j=1}^{s} a_{ij} \, f(U_j) + \Delta t \sum_{j=1}^{s} \bar{a}_{ij} \, g(U_j)
\end{align*}
where $U_i$ are the different stages ($i = 1, 2, \hdots, s$), $s$ is the total number of stages, and $U_s$ is the final solution at the $n^\mathrm{th}$ time step. The Butcher table of an ARK scheme can be represented by
\begin{equation*}
	\begin{array}{c|ccc|cccc}
		0   & 0  &&& 0 \\
		c_2 & a_{21}  & a_{22}  && \bar{a}_{21} & 0 \\
		\vdots & \vdots   & \ddots & & \vdots & \ddots & \ddots \\
		c_s = 1  & a_{s1}  & \hdots  & a_{ss} & \bar{a}_{s1} & \hdots & \bar{a}_{s \, s-1} & 0. \\
		\hline
	\end{array}
\end{equation*}
Similar to Ref. \cite{Crouseilles15}, we consider a second-order and a fourth-order integrator. The Butcher table of the 2-stage ARK.2.A.1 \citep{Liu06} integrator is
\begin{equation*}
	\begin{array}{c|ccc|ccc}
		0   & 0 && & 0 && \\
		1/2 & -1/2  & 1   &  & 1/2 &  &  \\
		1   & 1     & -1  & 1 & 0   & 1 & \\
		\hline
	\end{array}
\end{equation*}
and the set of equations can be expanded to
\begin{align*}
\begin{split}
	U_1 & = u^n \\
	U_2 & = u^n + \Delta t \left(-\frac{1}{2} f(U_1) + f(U_2) \right) + \frac{\Delta t}{2} g(U_1) \\
	U_3 & = u^n + \Delta t \left(f(U_1) - f(U_2) + f(U_3) \right) + \Delta t \, g(U_2).
\end{split}
\end{align*}
Here, we have $f(u) \equiv \lambda \nabla^2 u$ and $g(u) \equiv \nabla \cdot (\mathcal{D} \nabla u) - \lambda \nabla^2 u$:
\begin{align*}
\begin{split}
	U_1 & = u^n \\
	\left(I - \Delta t \lambda \nabla^2 \right) \, U_2 & = u^n + \frac{\Delta t}{2} \, \left(\nabla \cdot (\mathcal{D} \nabla U_1) - 2\lambda \nabla^2 U_1 \right) \\
	\left(I - \Delta t \lambda \nabla^2 \right) \, U_3 & = u^n + \Delta t \, \left(\nabla \cdot (\mathcal{D} \nabla U_2) - 2\lambda \nabla^2 U_2 + \lambda \nabla^2 U_1 \right) 
\end{split}
\end{align*}
The second-order solution of this scheme is given by $U_3$. The Butcher table of the 6-stage ARK.4.A.1 \citep{Liu06} integrator is
\begin{equation*}
	\begin{array}{c|ccccccc|cccccc}
		0 & 0 &&&&&&& 0 \\
		1/3 & -1/6 & 1/2 &&&&&& 1/3 \\
		1/3 & 1/6 & -1/3 & 1/2 &&&&& 1/6 & 1/6 \\
		1/2 & 3/8 & -3/8 & 0 & 1/2 &  &  &  & 1/8 & 0 & 3/8 &  &  & \\
		1/2 & 1/8 & 0 & 3/8 & -1/2 & 1/2 &  &  & 1/8 & 0 & 3/8 &  &  & \\
		1 & -1/2 & 0 & 3 & -2 & 0 & 1/2 &  & 1/2 & 0 & -3/2 & 1 & 1 & \\
		1 & 1/6 & 0 & 0 & 0 & 2/3 & -1/2 & 2/3 & 1/6 & 0 & 0 & 0 & 2/3 & 1/6 \\
		\hline
	\end{array}
\end{equation*}
which can be expanded to
\begin{align*}
\begin{split}
    U_1 & = u^n \\
    U_2 & = u^n + \Delta t \left(-\frac{1}{6} f(U_1) + \frac{1}{2}f(U_2) \right) + \frac{\Delta t}{3} g(U_1) \\
    U_3 & = u^n + \Delta t \left(\frac{1}{6}f(U_1) - \frac{1}{3}f(U_2) + \frac{1}{2}f(U_3) \right) + \Delta t \left(\frac{1}{6}g(U_1) + \frac{1}{6}g(U_2) \right) \\
    U_4 & = u^n + \Delta t \left(\frac{3}{8}f(U_1) - \frac{3}{8}f(U_2) + \frac{1}{2}f(U_4) \right) + \Delta t \left(\frac{1}{8}g(U_1) + \frac{3}{8}g(U_3) \right) \\
    U_5 & = u^n + \Delta t \left(\frac{1}{8}f(U_1) + \frac{3}{8}f(U_3) - \frac{1}{2}f(U_4) + \frac{1}{2}f(U_5) \right) + \Delta t \left(\frac{1}{8}g(U_1) + \frac{3}{8}g(U_3) \right) \\
    U_6 & = u^n + \Delta t \left(-\frac{1}{2}f(U_1) + 3f(U_3) - 2f(U_4) + \frac{1}{2}f(U_6) \right) + \Delta t \left(\frac{1}{2}g(U_1) - \frac{3}{2}g(U_3) + g(U_4) + g(U_5) \right) \\
    U_7 & = u^n + \Delta t \left(\frac{1}{6}f(U_1) + \frac{2}{3}f(U_5) - \frac{1}{2}f(U_6) + \frac{2}{3}f(U_7) \right) + \Delta t \left(\frac{1}{6}g(U_1) + \frac{2}{3}g(U_5) + \frac{1}{6}g(U_6) \right).
\end{split}
\end{align*}
Inserting the values of $f(u)$ and $g(u)$, we get
\begin{align*}
\begin{split}
    U_1 & = u^n \\
    \left(I - \frac{\Delta t}{2} \lambda \nabla^2 \right) \, U_2 & = u^n + \Delta t \left(-\frac{1}{2}\lambda \nabla^2 U_1 + \frac{1}{3}\nabla \cdot (\mathcal{D} \nabla U_1) \right) \\
    \left(I - \frac{\Delta t}{2} \lambda \nabla^2 \right) \, U_3 & = u^n + \Delta t \left(-\frac{1}{2}\lambda \nabla^2 U_2 + \frac{1}{6}\nabla \cdot (\mathcal{D} \nabla U_1) + \frac{1}{6}\nabla \cdot (\mathcal{D} \nabla U_2) \right) \\
    \left(I - \frac{\Delta t}{2} \lambda \nabla^2 \right) \, U_4 & = u^n + \Delta t \left(\frac{1}{4}\lambda \nabla^2 U_1 - \frac{3}{8}\lambda \nabla^2 U_2 - \frac{3}{8}\lambda \nabla^2 U_3 + \frac{1}{8}\nabla \cdot (\mathcal{D} \nabla U_1) + \frac{3}{8}\nabla \cdot (\mathcal{D} \nabla U_3) \right) \\
    \left(I - \frac{\Delta t}{2} \lambda \nabla^2 \right) \, U_5 & = u^n + \Delta t \left(-\frac{1}{2}\lambda \nabla^2 U_4 + \frac{1}{8}\nabla \cdot (\mathcal{D} \nabla U_1) + \frac{3}{8}\nabla \cdot (\mathcal{D} \nabla U_3) \right) \\
    \left(I - \frac{\Delta t}{2} \lambda \nabla^2 \right) \, U_6 & = u^n + \Delta t \left(-\lambda \nabla^2 U_1 + \frac{9}{2}\lambda \nabla^2 U_3 - 3\lambda \nabla^2 U_4 - \lambda \nabla^2 U_5\right) \\ 
    & + \Delta t \left(\frac{1}{2}\nabla \cdot (\mathcal{D} \nabla U_1) - \frac{3}{2}\nabla \cdot (\mathcal{D} \nabla U_3) + \nabla \cdot (\mathcal{D} \nabla U_4) + \nabla \cdot (\mathcal{D} \nabla U_5) \right) \\
    \left(I - \frac{2}{3} \Delta t \lambda \nabla^2 \right) \, U_7 & = u^n + \Delta t \left(-\frac{2}{3}\lambda \nabla^2 U_6 + \frac{1}{6}\nabla \cdot (\mathcal{D} \nabla U_1) + \frac{2}{3}\nabla \cdot (\mathcal{D} \nabla U_5) + \frac{1}{6}\nabla \cdot (\mathcal{D} \nabla U_6) \right).
\end{split}
\end{align*}
The fourth-order solution is given by $U_7$.


\subsection{Computation of the matrix exponential using Leja interpolation}

Since Eq. \eqref{eq:anidiff} is a linear equation, one can compute the \textit{exact} solution (subject to the error incurred in spatial discretisation) as $u^{n+1} = \exp(\nabla \cdot (\mathcal{D} \nabla) \Delta t)u^n$. In principle, one can choose an arbitrarily large value of $\Delta t$. However, in certain scenarios, this may run into some numerical instabilities. We do not encounter such issues for the problems considered in this work. We show that we can choose $\Delta t = T_f$, where $T_f$ is the final simulation time assuming that we start our simulations from $T = 0$, and we obtain excellent results in doing so (Sec. \ref{sec:results}). We note that computing the matrix exponential of large matrices using direct methods are computationally prohibitive.

The method of polynomial interpolation at Leja points \cite{Leja1957, Baglama98} was proposed to compute the matrix exponential and exponential-like functions appearing in exponential integrators \cite{Caliari04, Bergamaschi06, Caliari07b, Caliari09}. The focal point of this iterative scheme is to interpolate the action of the exponential of a matrix applied to a vector bounded by the spectrum of the underlying matrix. One needs an explicit estimate of the spectrum of the matrix, here, $A_\mathrm{diff} = (I_y \otimes A_{xx}) + (A_{yy} \otimes I_x) + (A_x \otimes A_y) + (A_y \otimes A_x)$. In this work, we do not consider time-dependent diffusion coefficients, and consequently $A_\mathrm{diff}$ remains constant in time. In such a case, one can compute the most dominant eigenvalue ($\beta$) of this matrix using the Gershgorin's disk theorem \cite{Gershgorin1931} for once and for all (i.e. it is valid for all times). To a good approximation, the smallest eigenvalue can be set to 0 \cite{Caliari14, Luan19, Deka22b}. We scale and shift the set of eigenvalues onto the the set of pre-computed Leja points ($\xi$) in the arbitrary spectral interval $[-2, 2]$. The scaling ($c$) and shifting ($\gamma$) factors are computed as $c = \beta/2$ and $\gamma = -\beta/4$, respectively. Now, we interpolate the function $\exp((c + \gamma \xi)\Delta t)$ on the set of Leja points, and the $(m+1)^\mathrm{th}$ term of the polynomial, so formed, is given by
\begin{align*}
    p_{m+1}(z) & = p_m(z) + d_{m+1} \, y_{m+1}(z), \\
    y_{m+1}(z) & = y_m(z) \times \left(\frac{z - c}{\gamma} - \xi_{m} \right),
\end{align*}
where $d_m$ are the coefficients of the polynomial determined using the divided differences algorithm \cite{deBoor2005}. Computation of $z = A_\mathrm{diff} y$, i.e. the RHS-function, constitutes the most expensive part of the Leja interpolation algorithm. The number of matrix-vector products computed is proportional to the number of Leja points needed for the polynomial to converge and can, thus, be used as a proxy of the computational cost. We use the publicly available \textsc{Python}-based \texttt{LeXInt} library \cite{Deka22_lexint} to compute the exponential of the matrix applied to a vector using the Leja interpolation scheme.


\subsection{Exponential Integrators and the \texorpdfstring{$\mu$}{TEXT}%
            -mode integrator}
            
Exponential integrators \cite{Ostermann10} are centered around the idea of writing an initial value problem of the form
\begin{equation*}
    \frac{\partial u}{\partial t} = f(u(t))
\end{equation*}
as 
\begin{equation}
    \frac{\partial u}{\partial t} = A u(t) + g(u(t)),
    \label{eq:ei}
\end{equation}            
where $A$ is the linear/stiff term and $g(u(t))$ is the nonlinear/non-stiff term. The first term on the RHS is solved exactly (in time) whereas $g(u(t))$ is treated using some explicit method. Notice that Eq. \eqref{eq:ei} has the same form as that of Eq. \eqref{eq:imex}. Once again, we penalise Eq. \eqref{eq:anidiff} by the Laplacian operator with a constant diffusion coefficient. We treat this constant coefficient diffusion term with the $\mu$-mode integrator \cite{Caliari22a, Caliari22b} and the remainder using the Leja interpolation method. Let us present a brief overview of the $\mu$-mode integrator. The diffusion equation reads
\begin{equation*}
    \frac{\partial u}{\partial t} = \frac{\partial ^2u}{\partial x^2} + \frac{\partial ^2u}{\partial y^2}
\end{equation*}
subject to the relevant initial and boundary conditions. Discretising this equation in space, one obtains 
\begin{equation}
    \frac{\partial u}{dt} = (I_x \otimes A_{yy} + A_{xx} \otimes I_y) u,
    \label{eq:kron_diff}
\end{equation}
where $A_{xx}$ and $A_{yy}$ are the Laplacian matrices obtained from the discretisation of $\frac{\partial ^2u}{\partial x^2}$ and $\frac{\partial ^2u}{\partial y^2}$, respectively. The solution of this equation at the $(n+1)^\mathrm{th}$ time step is given by
\begin{equation*}
    u^{n + 1} = \exp(\Delta t (I_x \otimes A_{yy} + A_{xx} \otimes I_y)) u^n.
\end{equation*}
Now, let us reformulate the action of the large matrices, i.e., $I_x \otimes A_{yy}$ and $A_{xx} \otimes I_y$, into the tensor notation: let $U(t)$ be a tensor of size $N_x \times N_y$, where $N_x$ and $N_y$ are the number of discretisation points along the X- and Y-directions, respectively. The columns (or rows) of $U(t)$ form the vector $u(t)$. This can easily be achieved by using the \texttt{numpy.meshgrid} command in \textsc{Python} or the \texttt{ndgrid} command in \textsc{Matlab}.  Eq. \eqref{eq:kron_diff} can now be rewritten as
\begin{equation*}
    \frac{\partial U}{dt} = A_{xx} U(t) + U(t) A_{yy}^{T},
\end{equation*}
the solution of which reads
\begin{equation}
    U^{n + 1} = \exp(\Delta t A_{xx}) U^n \exp(\Delta t A_{yy}^T) \footnote{This can also be written as $\exp((\Delta t A_{yy})(\exp(\Delta t A_{xx}) U^n)^T)^T$.}.
    \label{eq:mu_solution}
\end{equation}
Here, we have made use of the commutativity of $I_x \otimes A_{yy}$ and $A_{xx} \otimes I_y$ to obtain $\exp(\Delta t(I_y \otimes A_{xx})) = I_y \otimes \exp(\Delta t A_{xx})$ and $\exp(\Delta t(A_{yy} \otimes I_x)) = \exp(\Delta t A_{yy}) \otimes I_x$. This has been generalised to $d$ dimensions \cite{Caliari22a} and has recently been extended to $\varphi_l(z)$ functions \cite{Caliari22c}. The reformulation of the Kronecker product of the matrices into the tensor notation results in drastic computational savings. The exponential of the `smaller' (i.e. 1D) $A_{xx}$ and $A_{yy}$ matrices can be computed using standard methods such as the Pad\'e approximation or scaling and squaring that yields the exact solution.

In the context of the anisotropic diffusion equation with non-zero `drift' coefficients, i.e., $D_{xy}(x, y)$ and $D_{yx}(x, y)$, one cannot express the diffusion equation purely in terms of Kronecker sums. We obtain a remainder that consists of the term $(A_x \otimes A_y) + (A_y \otimes A_x)$. As such, one cannot directly apply the $\mu$-mode integrator to the anisotropic diffusion problem. This is why we resort to exponential integrators where we solve the constant coefficient (stiff) term using the $\mu$-mode integrator. We consider the second-order exponential time differencing Runge--Kutta (ETDRK2, Eq. \eqref{eq:etdrk2}) integrator \cite{Cox02}, the equations of which read as follows:
\begin{align}
    a^n = \exp(A \Delta t)u^n + \varphi_1(A \Delta t) g(u^n) \Delta t \nonumber \\
    u^{n+1} = a^n + \varphi_2(A \Delta t) (g(a^n) - g(u^n)) \Delta t
    \label{eq:etdrk2}
\end{align}
The $\varphi_l(z)$ functions applied to some vector, $g$, are evaluated using the Leja interpolation scheme. The algorithm is similar to the one described in the previous subsection.


\section{Test Problems}
\label{sec:test}

In this section, we describe the different two-dimensional test models considered in this study, in all of which we consider periodic boundary conditions in the spatial domain $X \times Y \in [-1, 1] \times [-1, 1]$.


\subsection*{Case I: Periodic Band}

\begin{figure}
    \centering
	\includegraphics[width = 0.965\textwidth]{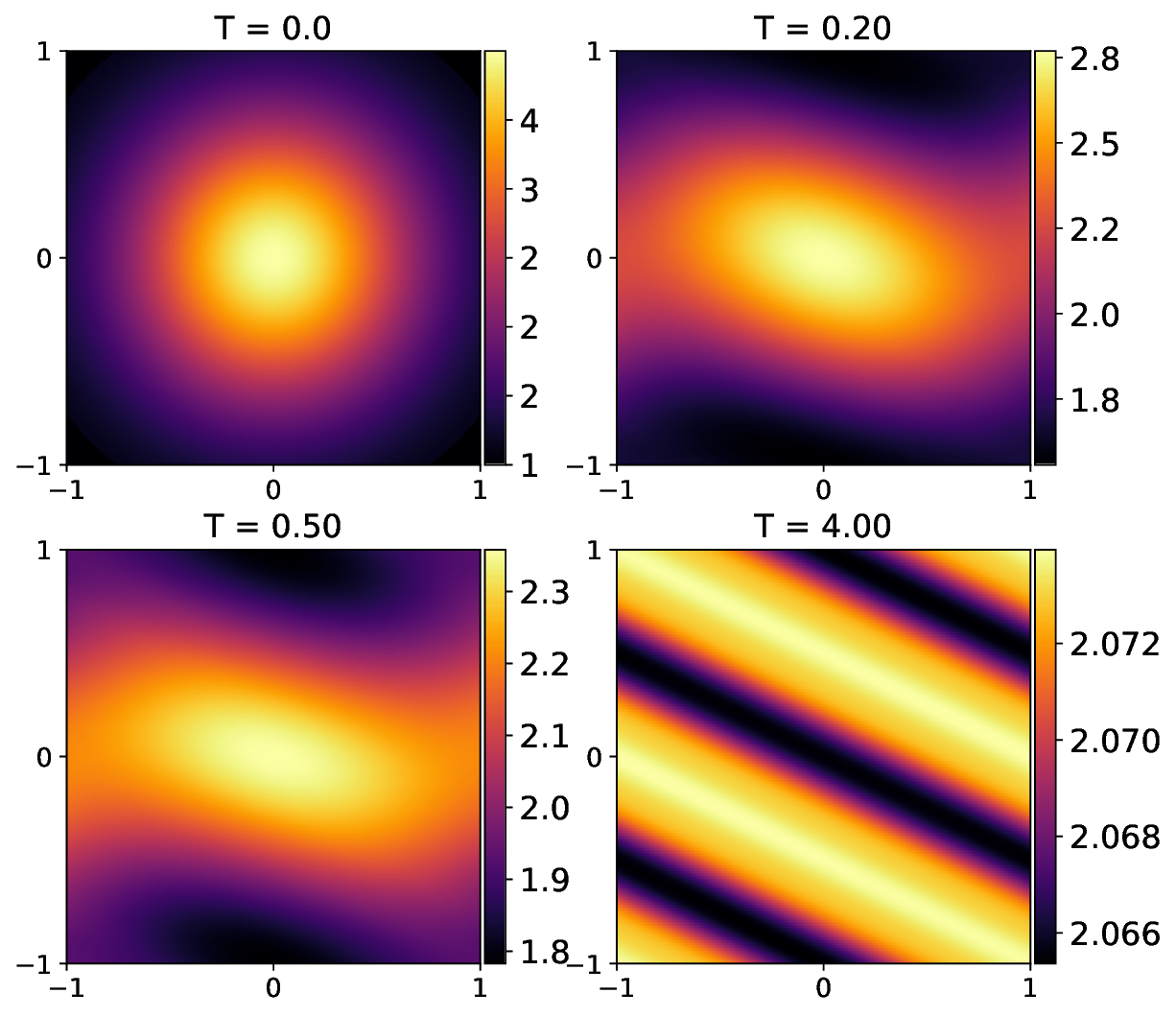}
	\includegraphics[width = 0.425\textwidth]{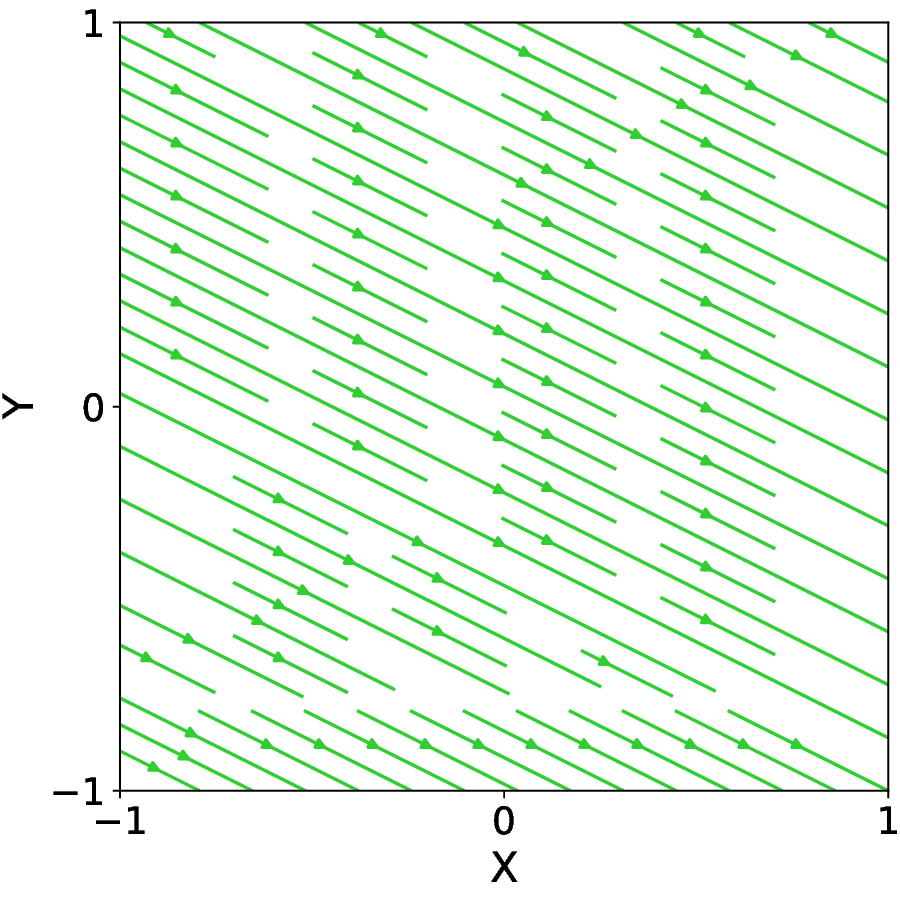}
    \caption{The distribution function, for the periodic-band (case I), is illustrated at four different times. At the bottom, we show the magnetic field configuration leading to the related diffusion tensor (Eq. \eqref{eq:B_band}).}
    \label{fig:time_evol_band}
\end{figure}

We start off with the diffusion of the Gaussian 
\begin{equation*}
    u(x, y, t = 0) = 
    \begin{cases}
        1 + 3\exp{(-2r^2)}  & \text{if $r = \sqrt{x^2 + y^2} < 2\pi/5$}, \\
        1                   & \text{otherwise},
    \end{cases}
    \label{eq:gaussian1}
\end{equation*}
as a periodic band (example taken from Ref. \cite{Crouseilles15}) where the coefficients of the diffusion tensor are given by
\begin{equation}
    \mathcal{D} = \begin{bmatrix}
                    1.00 & &    0.50 \\
                    0.50 & &    0.25 \\
        \end{bmatrix}.
    \label{eq:B_band}
\end{equation}
Note that the magnitude of the coefficients are an indication of the relative amounts of diffusion along the different directions. The diffusion tensor, which is characterised by the ambient magnetic field, is chosen to be constant in space and time. We have chosen three different final simulation times ($T_f$) to study the performance of the relevant temporal integrators during different evolution regimes: $T_f = 0.2, 0.5,$ and $4.0$. $T_f = 4.0$ corresponds to the steady-state solution. Fig. \ref{fig:time_evol_band} illustrates the distribution function at these three different evolution times, the initial state, and the magnetic field configuration. The steady state solution is expected to align with the magnetic field lines.

\subsection*{Case II: Gaussian Pulse}

\begin{figure}
    \centering
	\includegraphics[width = 0.965\textwidth]{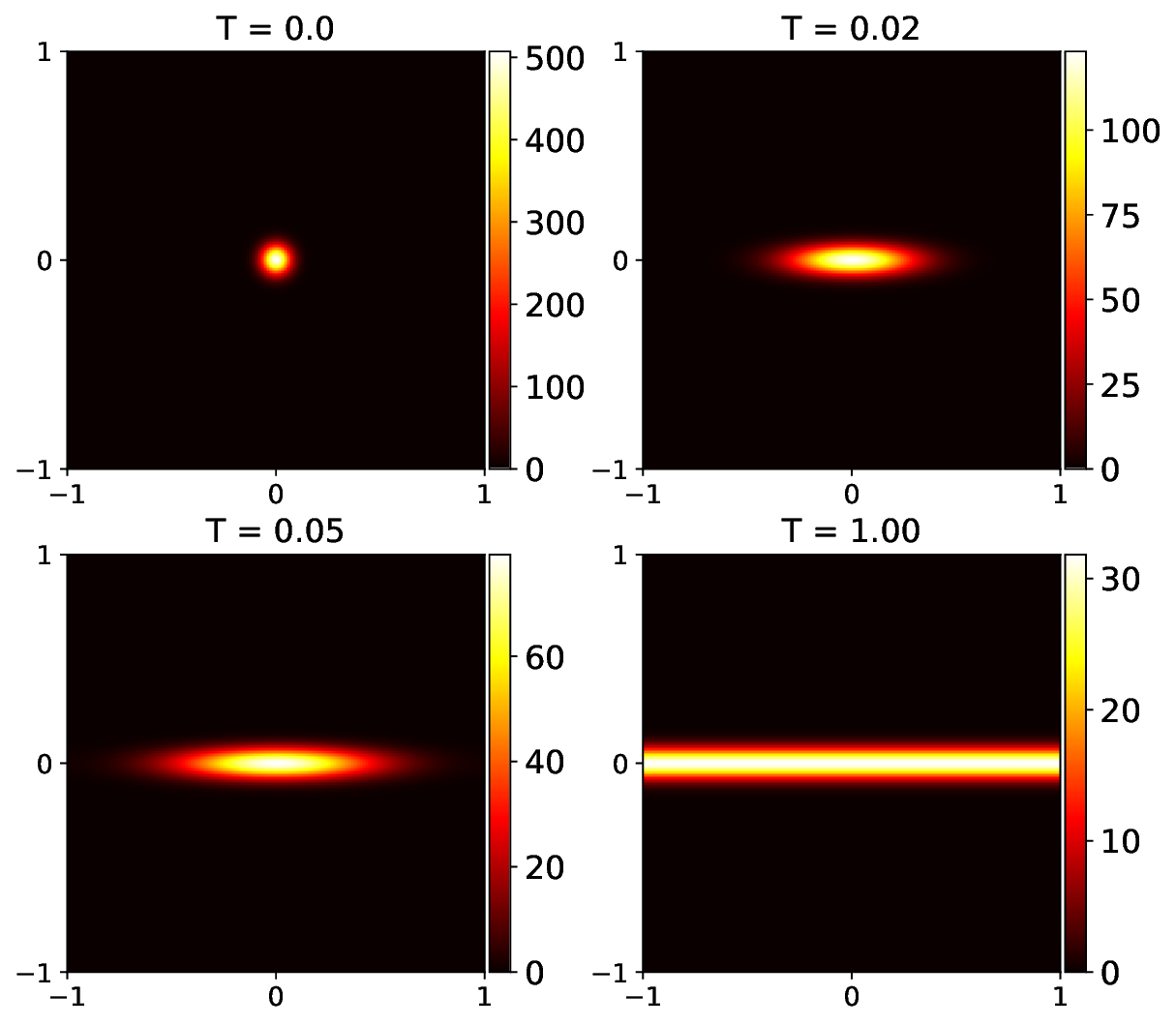}
	\includegraphics[width = 0.425\textwidth]{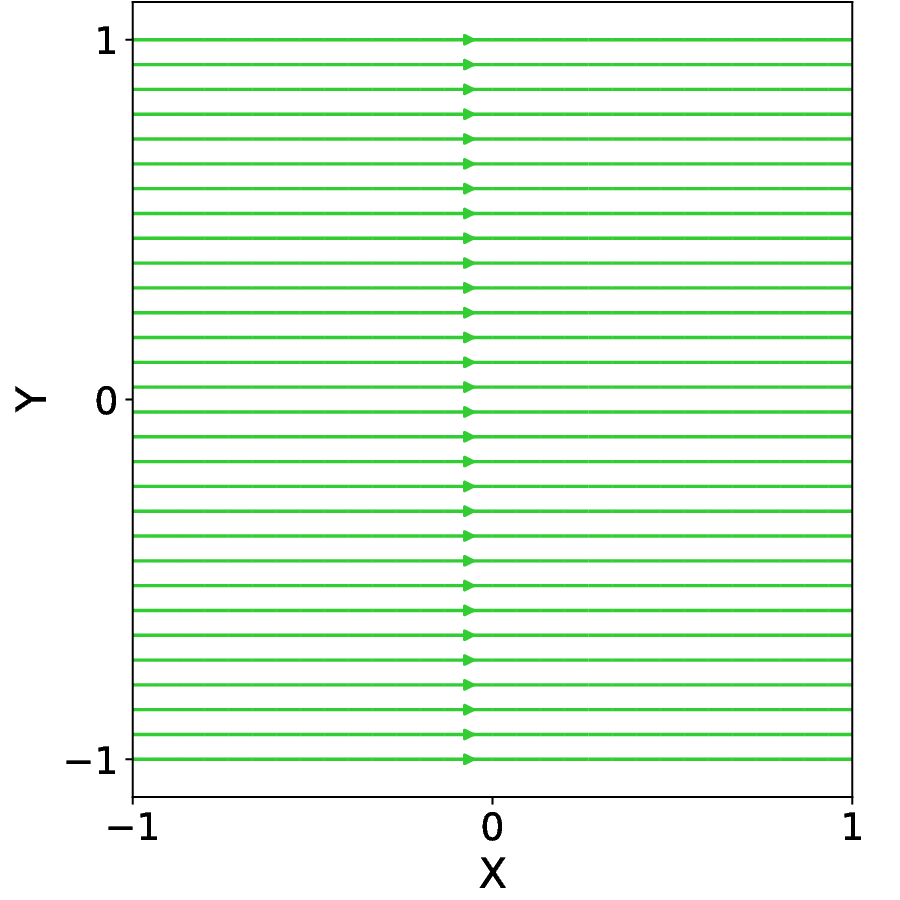}
    \caption{Same as Fig. \ref{fig:time_evol_band}, but for the Gaussian pulse (case II).}
    \label{fig:time_evol_gaussian}
\end{figure}

The next example, drawn from Ref. \cite{Hopkins17}, is the diffusion of the Gaussian pulse
\begin{equation*}
    u(x, y, t = 0) = \frac{(2\pi)^{-3/2}}{\epsilon^3} \exp\left[-\frac{1}{2}\left(\frac{x^2 + y^2}{\epsilon^2}\right)\right], \quad \epsilon = 0.05,
    \label{eq:gaussian2}
\end{equation*} 
along the X-axis. The diffusion matrix, 
\begin{equation}
    \mathcal{D} = \begin{bmatrix}
                    1 &     0 \\
                    0 &     0 \\
        \end{bmatrix},
    \label{eq:B_gaussian}
\end{equation}
is constant in space and time. There is no explicit diffusion along the Y-direction. The time-evolution of this Gaussian, at $T_f = 0.02, 0.05$, and $1.0$, and the magnetic field configuration are illustrated in Fig. \ref{fig:time_evol_gaussian}. Here, the mixed derivatives terms vanish, which allows us to directly apply the $\mu$-mode integrator. This test problem can be used a proxy for cases where there is small amount of diffusion in the perpendicular direction. The artificial diffusion, obtained as a byproduct due to the (second-order) spatial discretisation, is likely to introduce more diffusion than having explicit perpendicular diffusion that is roughly $4-5$ orders of magnitude smaller than the diffusion along the magnetic field lines.  

\subsection*{Case III: Ring}

\begin{figure}
    \centering
	\includegraphics[width = 0.95\textwidth]{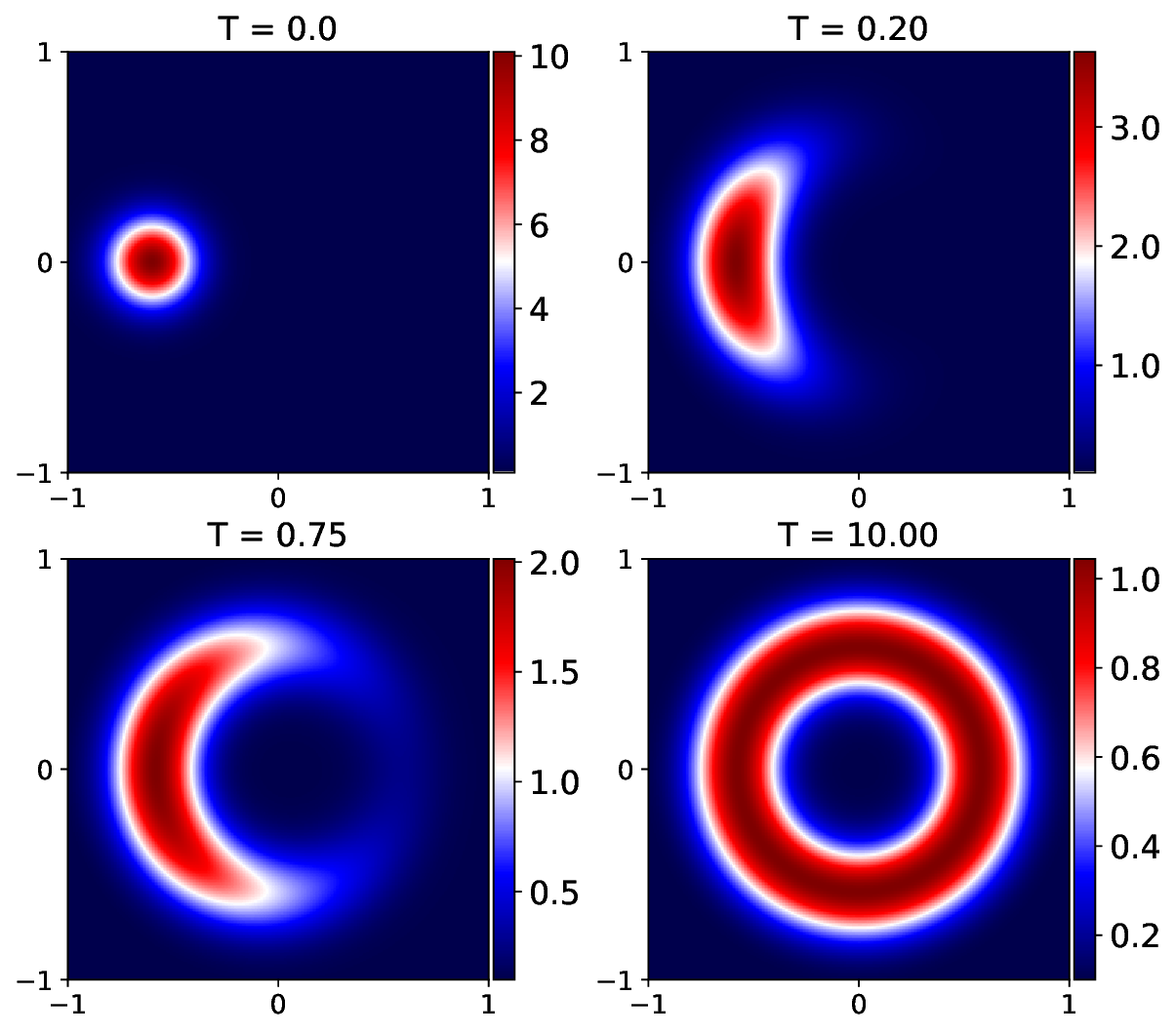}
	\includegraphics[width = 0.47\textwidth]{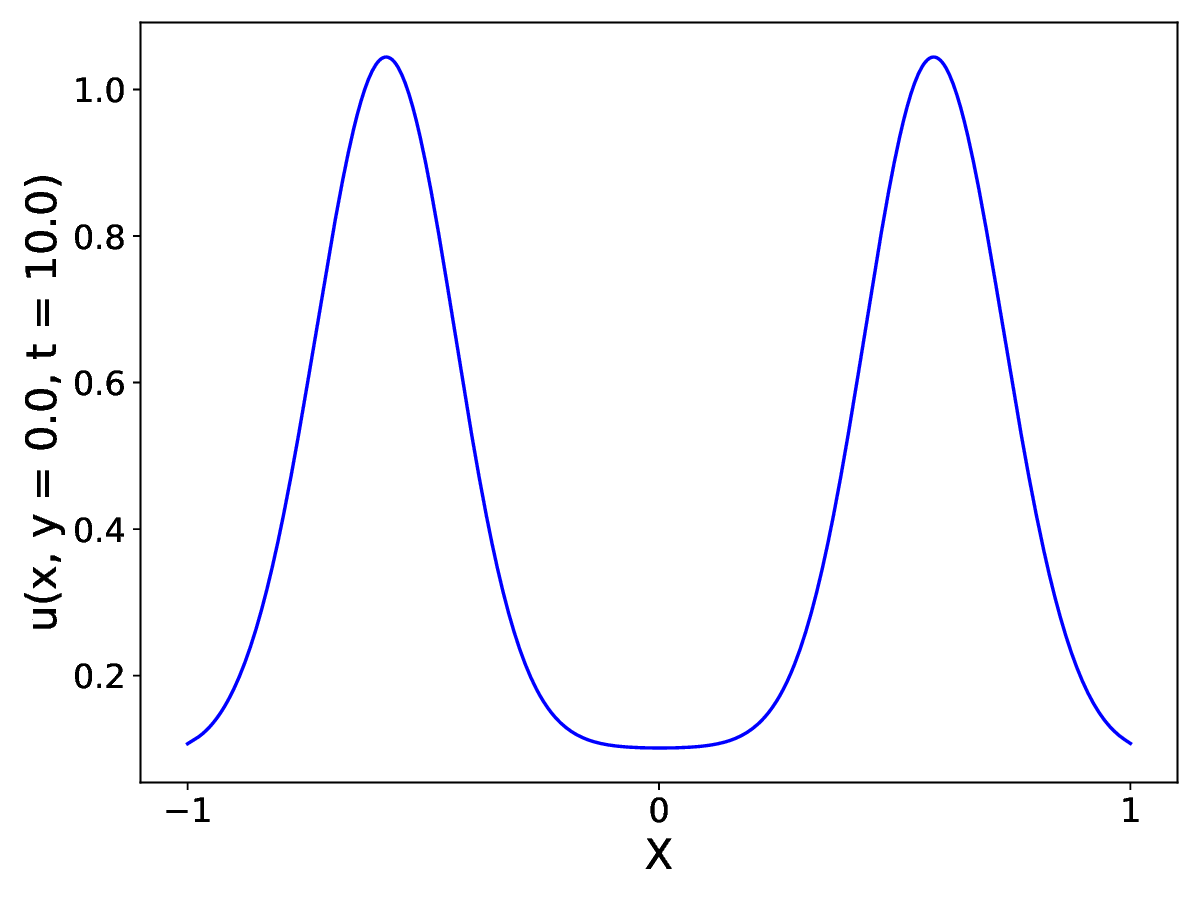}
	\includegraphics[width = 0.42\textwidth]{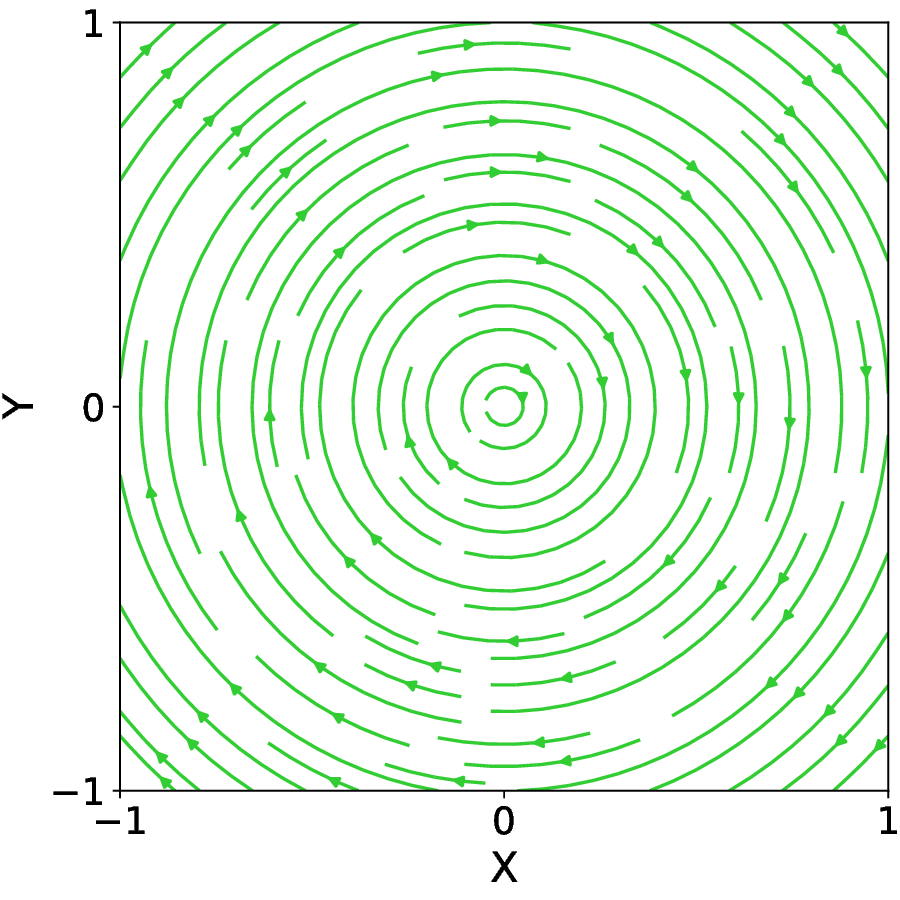}
    \caption{Same as Fig. \ref{fig:time_evol_band}, but for the ring (case III). Additionally, at the bottom left, we show the cross-section, along the $X-$direction, at $Y = 0.0$ and $T_f = 10.0$.}
    \label{fig:time_evol_ring}
\end{figure}

The final example is the diffusion on a ring \cite{Sharma07, Crouseilles15, Pakmor16b, Hopkins17}, where the diffusion coefficients depend on the spatial coordinates as follows:
\begin{equation}
    \mathcal{D} = \begin{bmatrix}
                     y^2 &    -xy   \\
                    -xy  &     x^2  \\
        \end{bmatrix}.
    \label{eq:B_ring}
\end{equation}
This is, obviously, a more realistic scenario for CR diffusion across the Galactic magnetic field. We choose the initial distribution to be
\begin{equation*}
    u(x, y, t = 0) = 0.1 + 10\exp\left(-\frac{(x + 0.6)^2 + y^2}{0.04}\right).
    \label{eq:gaussian3}
\end{equation*} 
As the solution evolves in time, the particles diffuse along concentric circles with no explicit diffusion in the radial direction (Fig. \ref{fig:time_evol_ring}). We choose three different simulation times $T_f = 0.2, 0.75,$ and $10.0$.  The steady state solution is characterised by a ring or an annular disc (owing to numerical diffusion in the perpendicular direction).

\section{Results}
\label{sec:results}


\subsection{Comparison of Leja with Crank--Nicolson}

\begin{figure}[t!]
    \centering
    \includegraphics[width = \textwidth]{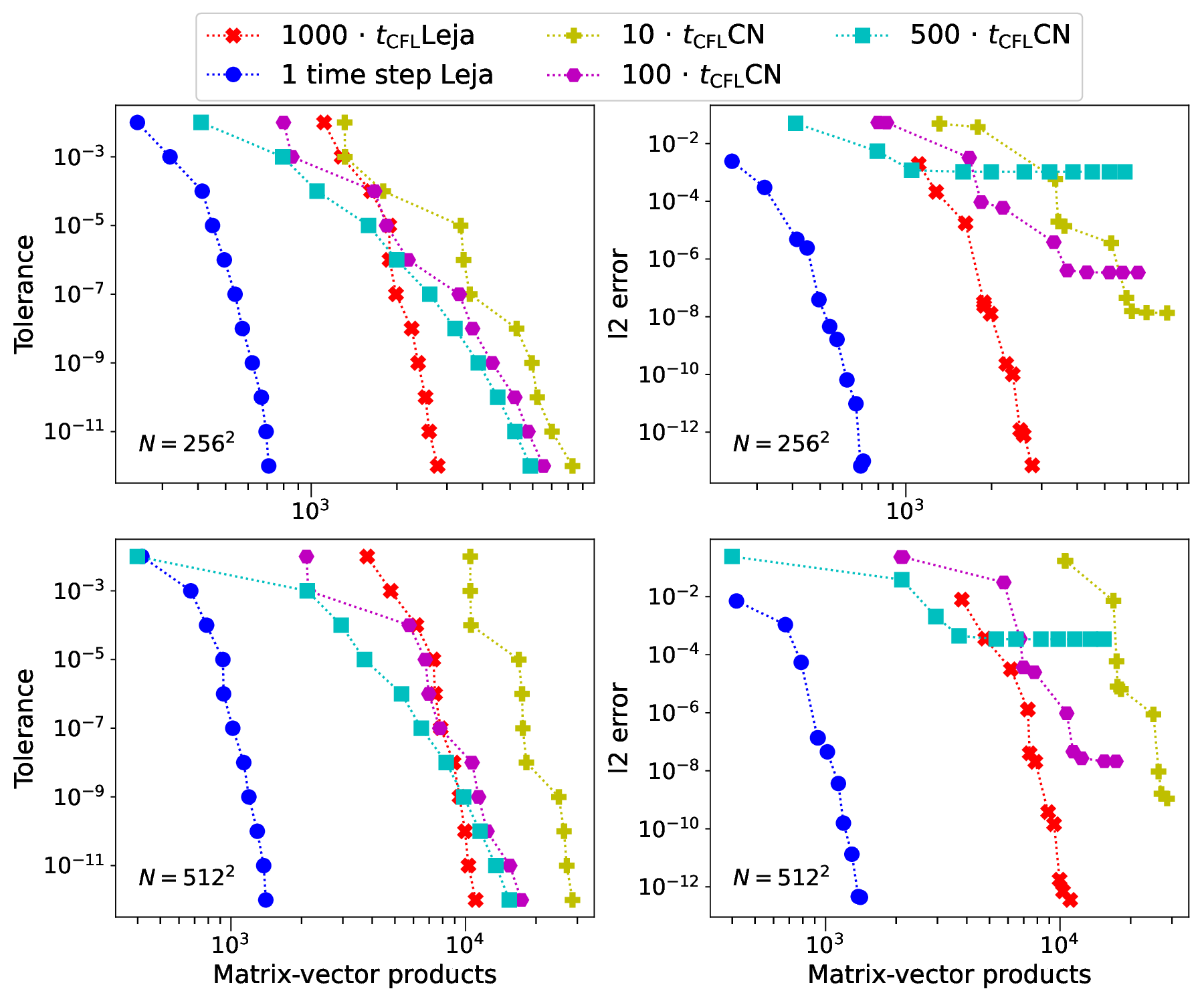}
    \caption{Comparison of the performance of Crank--Nicolson with the computation of the matrix exponential using Leja interpolation for Case I at T = 0.2. On the X-axis, we have the the number of matrix-vector products as a proxy for the computational cost. The left panel shows the user-defined tolerance whereas the right panel shows the l2 norm of the error incurred, as a function of the computational cost for two different spatial resolutions. Results are shown for different time-step sizes as shown in the legend.}
    \label{fig:cost_cn_b_0.2}
\end{figure}

\begin{figure}[t!]
    \centering
    \includegraphics[width = \textwidth]{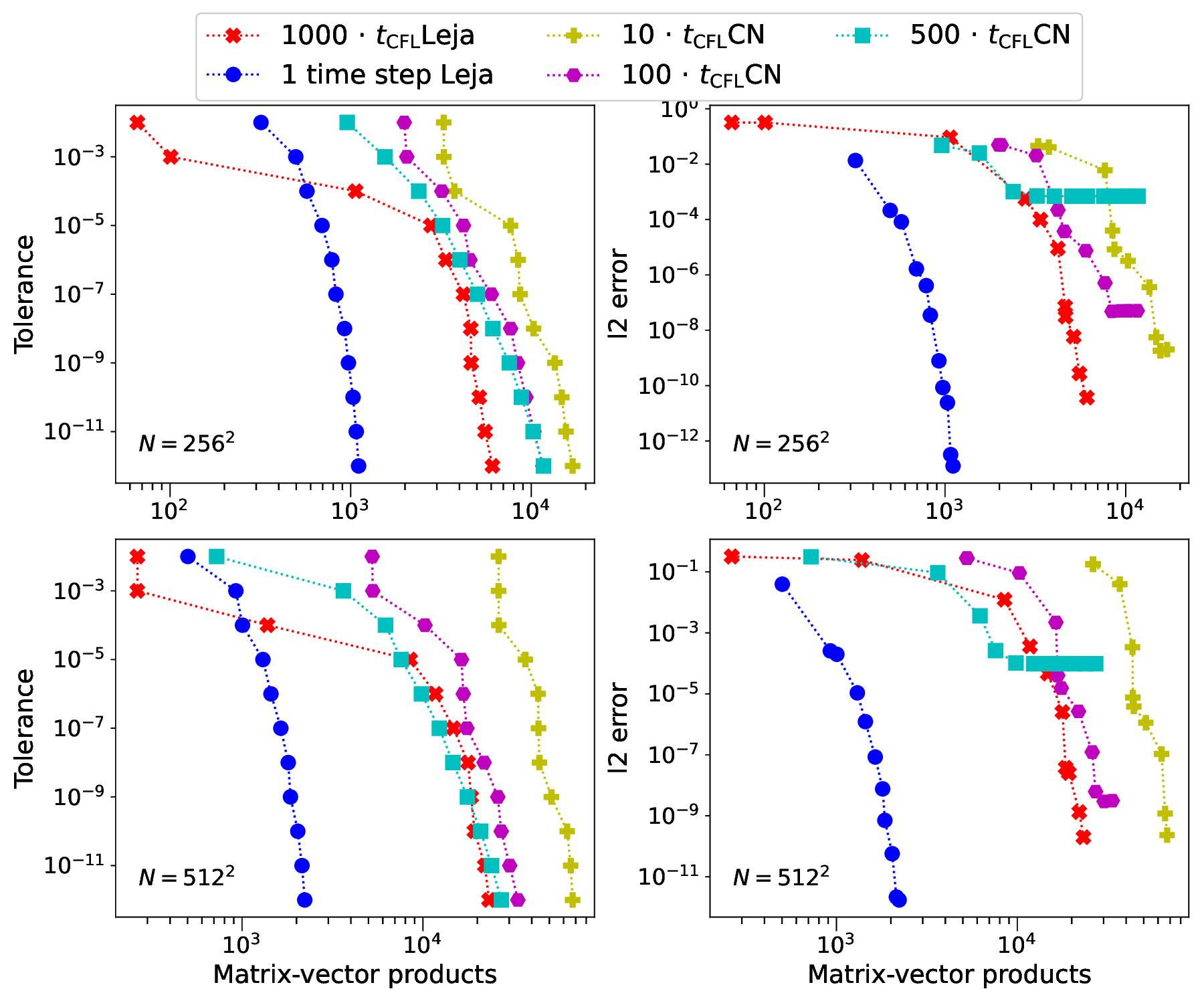}
    \caption{Same as Fig. \ref{fig:cost_cn_b_0.2}, but at T = 0.5.}
    \label{fig:cost_cn_b_0.5}
\end{figure}

\begin{figure}[t!]
    \centering
    \includegraphics[width = \textwidth]{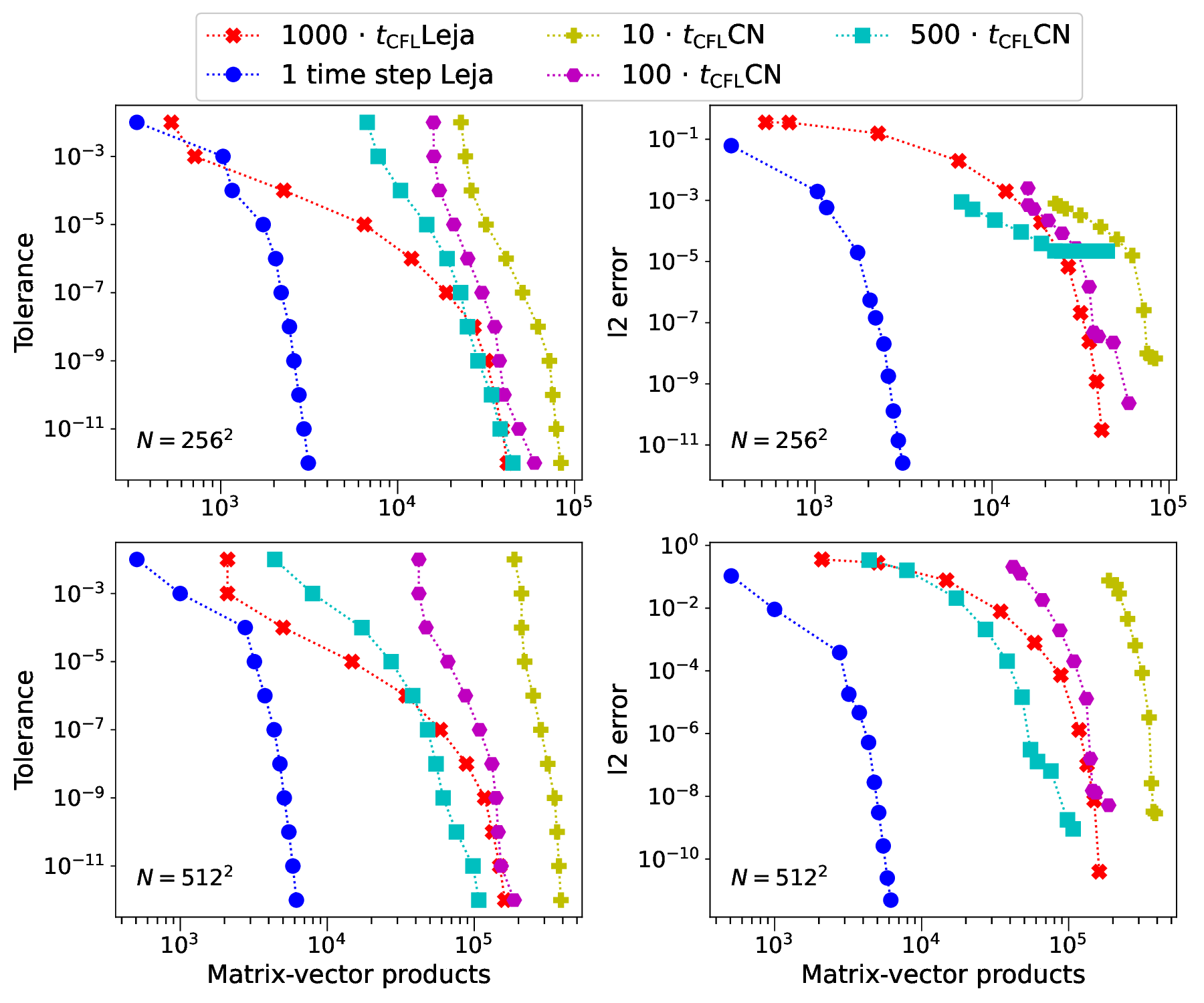}
    \caption{Same as Fig. \ref{fig:cost_cn_b_0.2}, but at T = 4.0 (steady-state solution).}
    \label{fig:cost_cn_b_4.0}
\end{figure}

Since Crank--Nicolson has been the traditionally favoured integrator for solving the time-dependent CR transport equation, we show the superiority in the performance of the Leja method that directly computes the matrix exponential over the Crank--Nicolson integrator. The results for the case of the diffusion in a periodic band are illustrated in Figs. \ref{fig:cost_cn_b_0.2}, \ref{fig:cost_cn_b_0.5}, and \ref{fig:cost_cn_b_4.0}. The most expensive part of solving the anisotropic diffusion equation, with the Leja as well as the Crank--Nicolson method, is the computation of the matrix-vector products, which we consider to be a proxy of the computational cost. Let us clearly state that we are exclusively interested in studying the performance of temporal integrators in this study. The error (incurred during the time integration of the anisotropic diffusion equation) is estimated by comparing the simulations to a reference solution computed using very small step sizes for a given spatial resolution. The l2 norm of the error incurred can be computed as 
\[ \mathrm{l2 \, error} = \frac{1}{||u_\mathrm{reference}||} \sum_{i=0}^{N_x \times N_y} (u_{\mathrm{solution}, i} - u_{\mathrm{reference}, i})^2. \] For simulation times of  $T = 0.2$ and $0.5$, i.e., the transient stage, we observe that Crank--Nicolson fails to yield accurate solutions (l2 error $\leq 10^{-5}$) for large step sizes, i.e. $\Delta t  = 500 \cdot t_\mathrm{CFL}$. Therefore, one is forced to choose smaller step sizes to generate better results, in terms of accuracy, thereby significantly upsurging the computational cost. The direct computation of the matrix exponential using the Leja method, however, is able to generate highly accurate solutions, depending on the user-defined tolerance, whilst taking substantially larger step sizes. As a matter of fact, one can choose to have only one time step, where the step size is equal to the final simulation time ($\Delta t = T_f$). This is clearly the best case scenario as one is able to reduce the number of matrix-vector products by almost two orders of magnitude without compromising the accuracy of the solution. We note that in some configurations, especially in cases with low spatial resolution and relatively small simulation times, the blue and the red markers coincide. This is due to the fact in these low resolution simulations, $1000 \cdot \Delta t_\mathrm{CFL} \geq T_f$, which is why we only have a single time step.

\begin{figure}[t!]
    \centering
    \includegraphics[width = \textwidth]{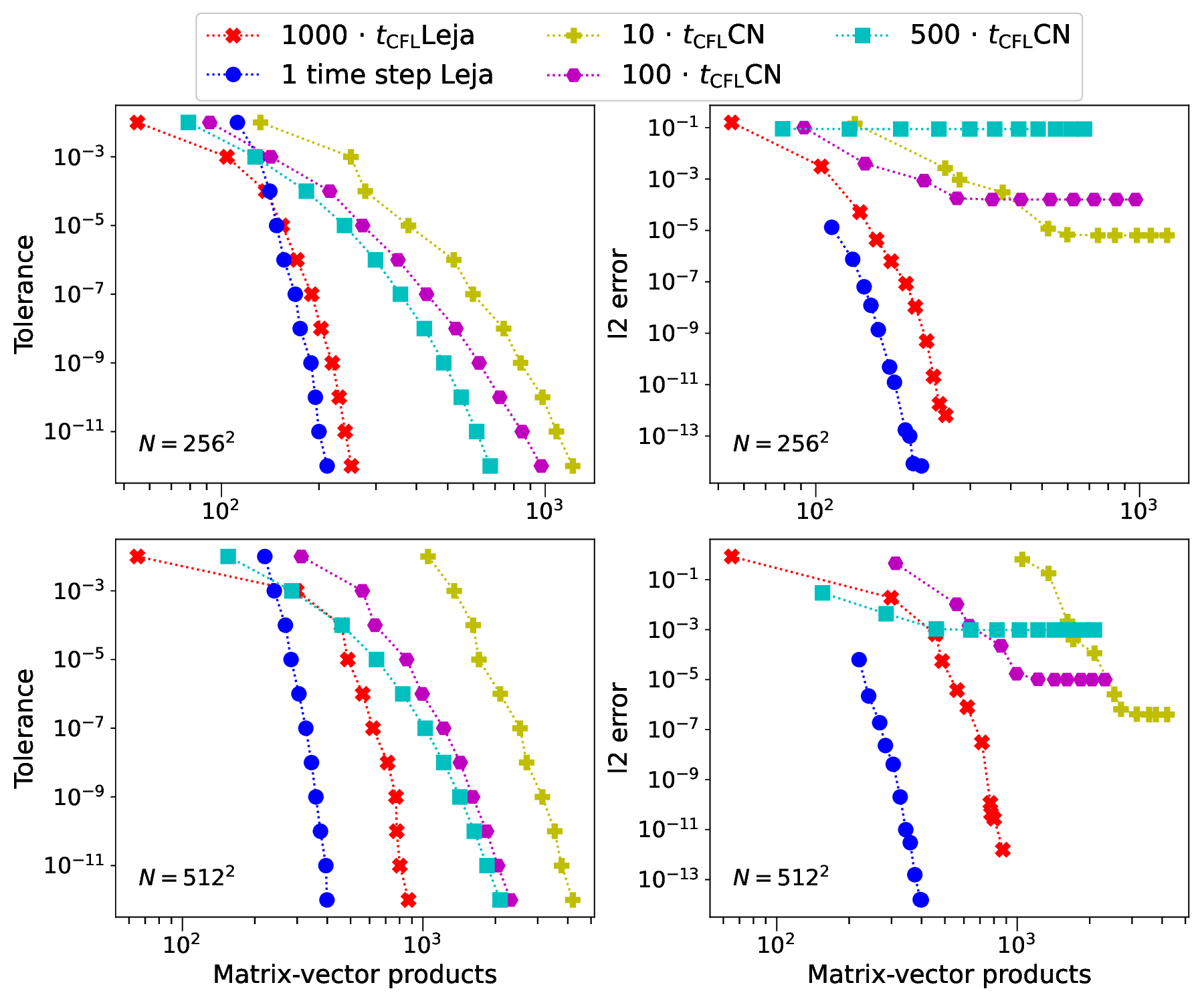}
    \caption{Comparison of the performance of Crank--Nicolson with the computation of the matrix exponential using Leja interpolation for Case II at T = 0.02. On the X-axis, we have the the number of matrix-vector products as a proxy for the computational cost. The left panel shows the user-defined tolerance whereas the right panel shows the l2 norm of the error incurred, as a function of the computational cost for two different spatial resolutions. Results are shown for different time-step sizes as shown in the legend.}
    \label{fig:cost_cn_g_0.02}
\end{figure}

\begin{figure}[t!]
    \centering
    \includegraphics[width = \textwidth]{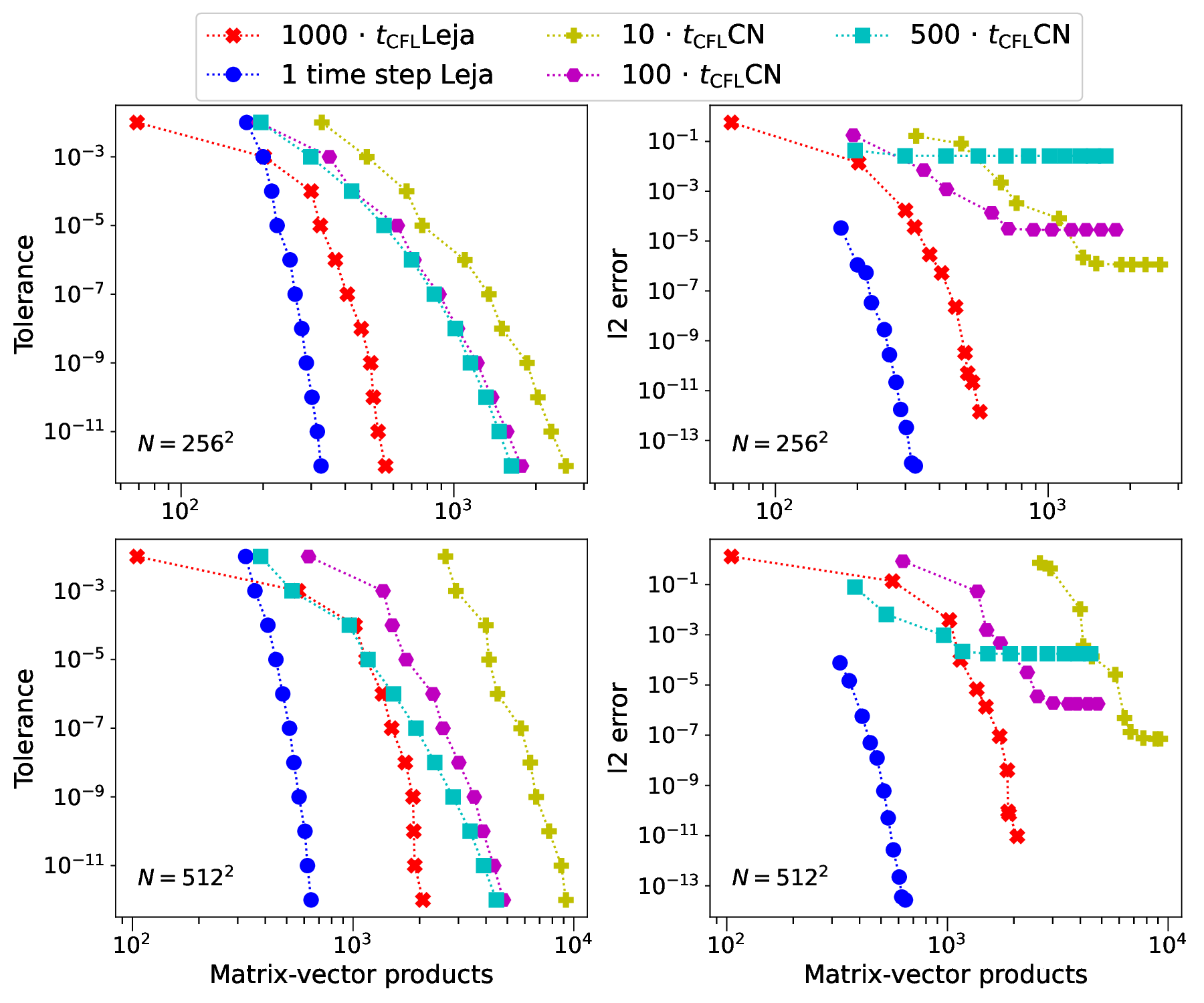}
    \caption{Same as Fig. \ref{fig:cost_cn_g_0.02}, but at T = 0.05.}
    \label{fig:cost_cn_g_0.05}
\end{figure}

\begin{figure}[t!]
    \centering
    \includegraphics[width = \textwidth]{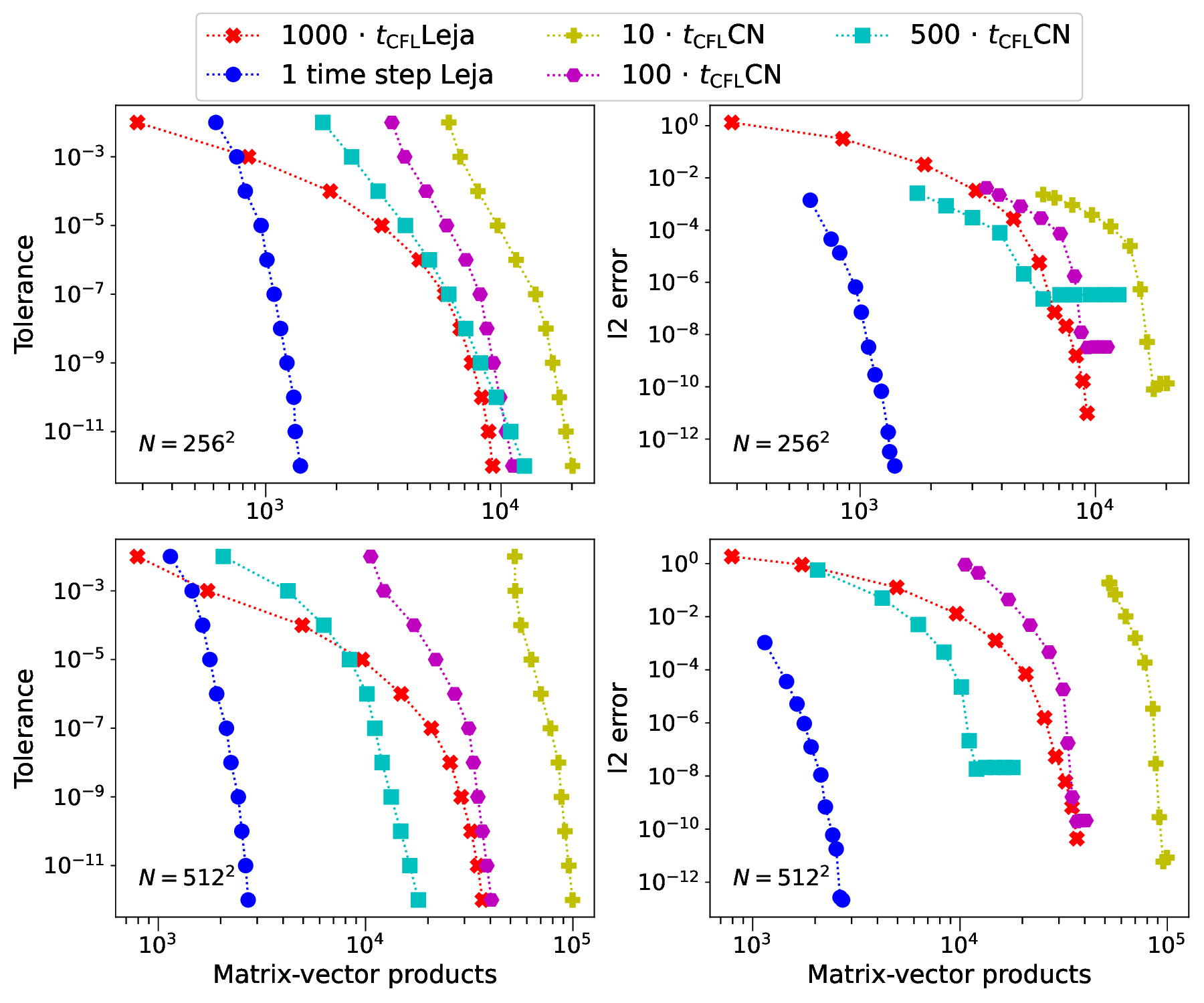}
    \caption{Same as Fig. \ref{fig:cost_cn_g_0.02}, but at T = 1.00 (steady-state solution).}
    \label{fig:cost_cn_g_1.0}
\end{figure}

\begin{figure}[t!]
    \centering
    \includegraphics[width = \textwidth]{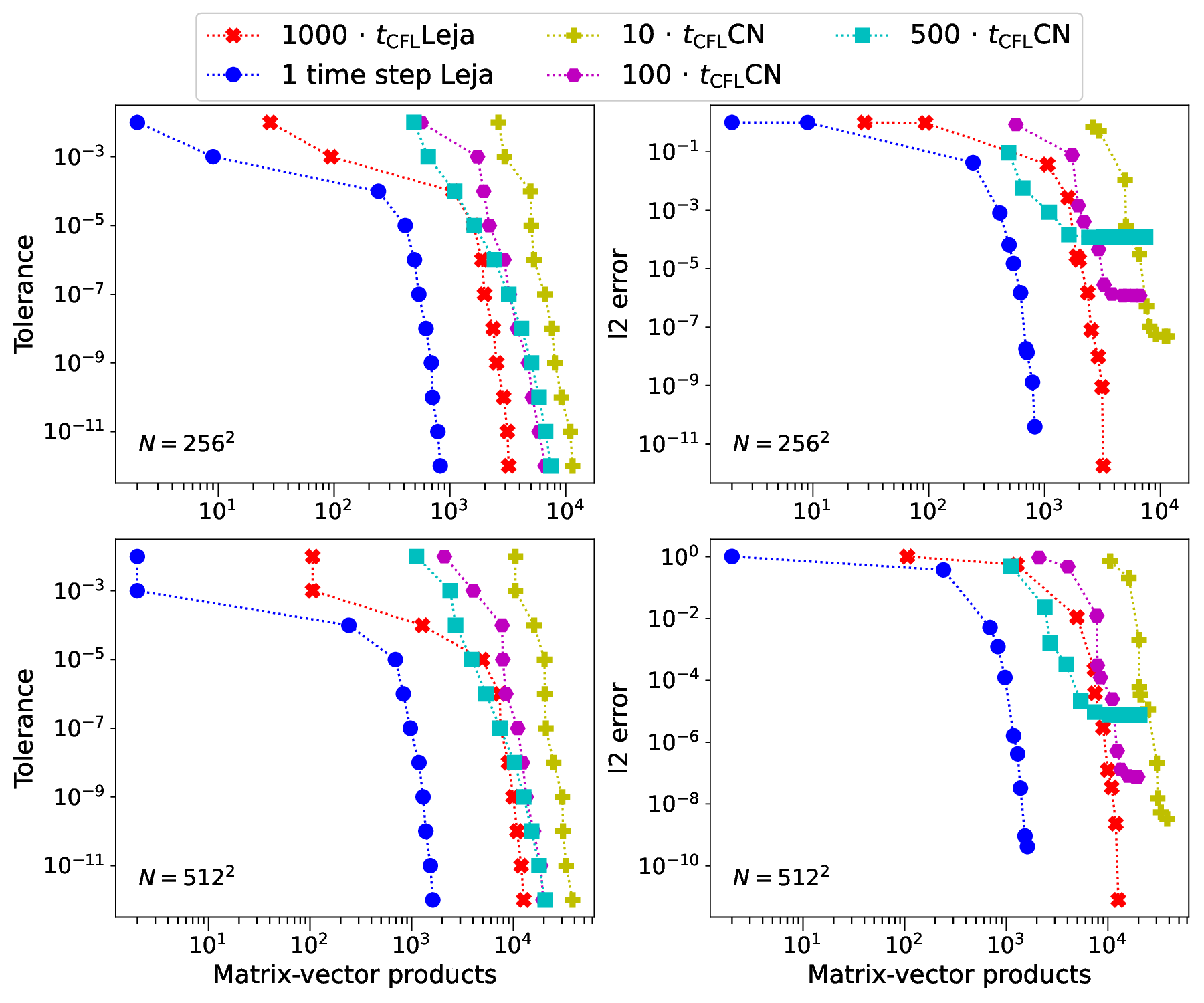}
    \caption{Comparison of the performance of Crank--Nicolson with the computation of the matrix exponential using Leja interpolation for Case III at T = 0.2. On the X-axis, we have the the number of matrix-vector products as a proxy for the computational cost. The left panel shows the user-defined tolerance whereas the right panel shows the l2 norm of the error incurred, as a function of the computational cost for two different spatial resolutions. Results are shown for different time-step sizes as shown in the legend.}
    \label{fig:cost_cn_r_0.2}
\end{figure}

\begin{figure}[t!]
    \centering
    \includegraphics[width = \textwidth]{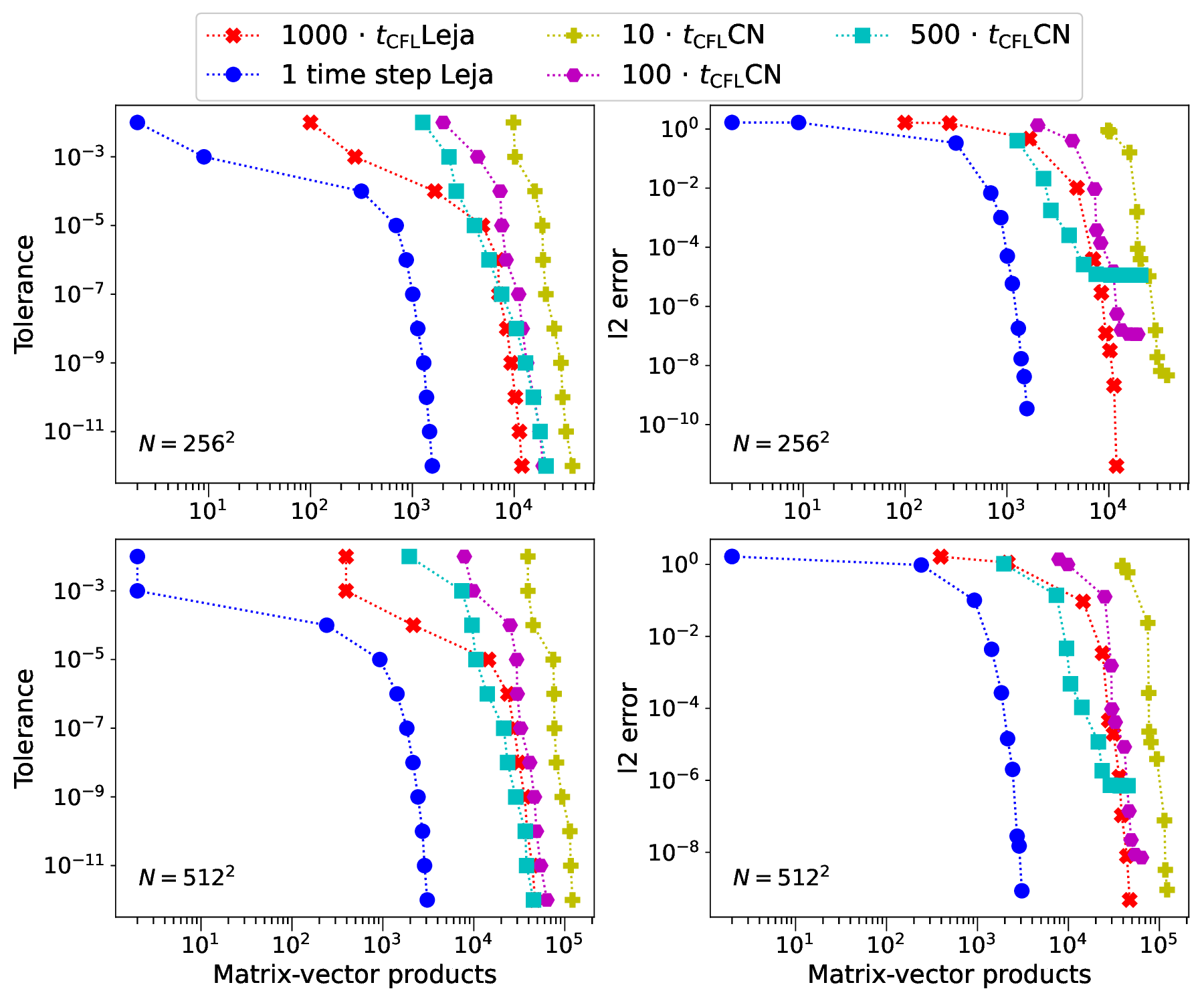}
    \caption{Same as Fig. \ref{fig:cost_cn_r_0.2}, but at T = 0.75.}
    \label{fig:cost_cn_r_0.75}
\end{figure}

\begin{figure}[t!]
    \centering
    \includegraphics[width = \textwidth]{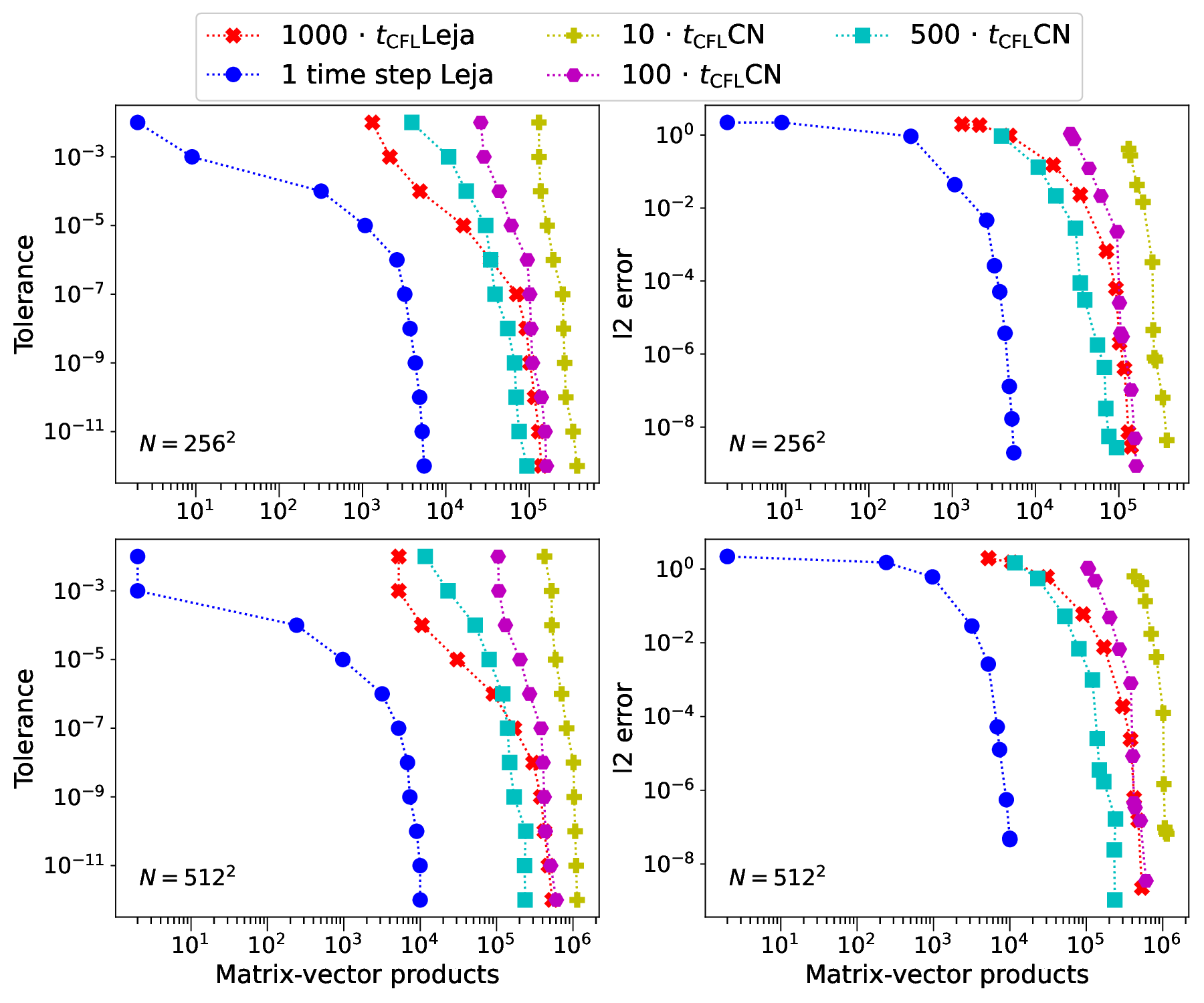}
    \caption{Same as Fig. \ref{fig:cost_cn_r_0.2}, but at T = 10.0 (steady-state solution).}
    \label{fig:cost_cn_r_10.0}
\end{figure}

As the solution approaches steady state, the Crank--Nicolson solver tends to become more accurate (Fig. \ref{fig:cost_cn_b_4.0}). The amplitude of the distribution function becomes minute, since most of it is smeared out as the solution approaches the steady state. Consequently, the error incurred decreases by a significant margin. We observe that the computational cost, for the Crank--Nicolson solver, plummets by a substantial margin without hampering the accuracy of the solution (Figs. \ref{fig:cost_cn_b_4.0}, \ref{fig:cost_cn_g_1.0}, and \ref{fig:cost_cn_r_10.0}). Nevertheless, it it still at least an order of magnitude more expensive than the Leja method (for $\Delta t = T_f$). It is to be noted that we are not always interested in the steady-state solution, but rather at some intermediate time in order to understand the physics of the problem under consideration. In the presence of sources, advection, or other processes, understanding the evolution of the distribution function over time, or at some transient time, may become more important than the steady-state solution. This is where the method of computing the matrix exponential by Leja interpolation becomes radically superior, both in terms of accuracy and cost, to the Crank--Nicolson solver (Figs. \ref{fig:cost_cn_b_0.2}, \ref{fig:cost_cn_b_0.5}, \ref{fig:cost_cn_g_0.02}, \ref{fig:cost_cn_g_0.05}, \ref{fig:cost_cn_r_0.2}, and \ref{fig:cost_cn_r_0.75}).

Similar observations have been made for the cases of diffusion of the Gaussian pulse (Figs. \ref{fig:cost_cn_g_0.02}, \ref{fig:cost_cn_g_0.05}, and \ref{fig:cost_cn_g_1.0}) and that on a ring (Figs. \ref{fig:cost_cn_r_0.2}, \ref{fig:cost_cn_r_0.75}, and \ref{fig:cost_cn_r_10.0}). For low resolution simulations, Crank--Nicolson fails to converge to a reasonable accuracy for almost all values of tolerance. The Leja interpolation method performs equal to, if not better than, Crank--Nicolson for $\Delta t = 1000 \cdot t_\mathrm{CFL}(< T_f)$ in terms of the computational cost. A further reduction in the cost can be obtained by choosing a larger step size, i.e. only one time step ($\Delta t = T_f$). For high-resolution simulations, this can result in savings up to almost two orders of magnitude.

We also note that the use of an efficient preconditioner may alleviate the need for an unnaturally large number of iterations to solve the systems of linear equations. For example, \textsc{GalProp} v57 implements the BiCGSTAB solver, making use of either the diagonal or IncompleteLUT pre-conditioner, with the Crank--Nicolson integrator \cite{Porter22}. Nevertheless, the accuracy of the solution, whilst using the Crank--Nicolson solver would still be limited to second order in time. No such restrictions apply to the Leja interpolation method.

\subsection{Performance of ARK methods, \texorpdfstring{$\mu$}{TEXT}%
            -mode integrator, and Exponential Integrators}

\begin{figure}[t!]
    \centering
    \includegraphics[width = \textwidth]{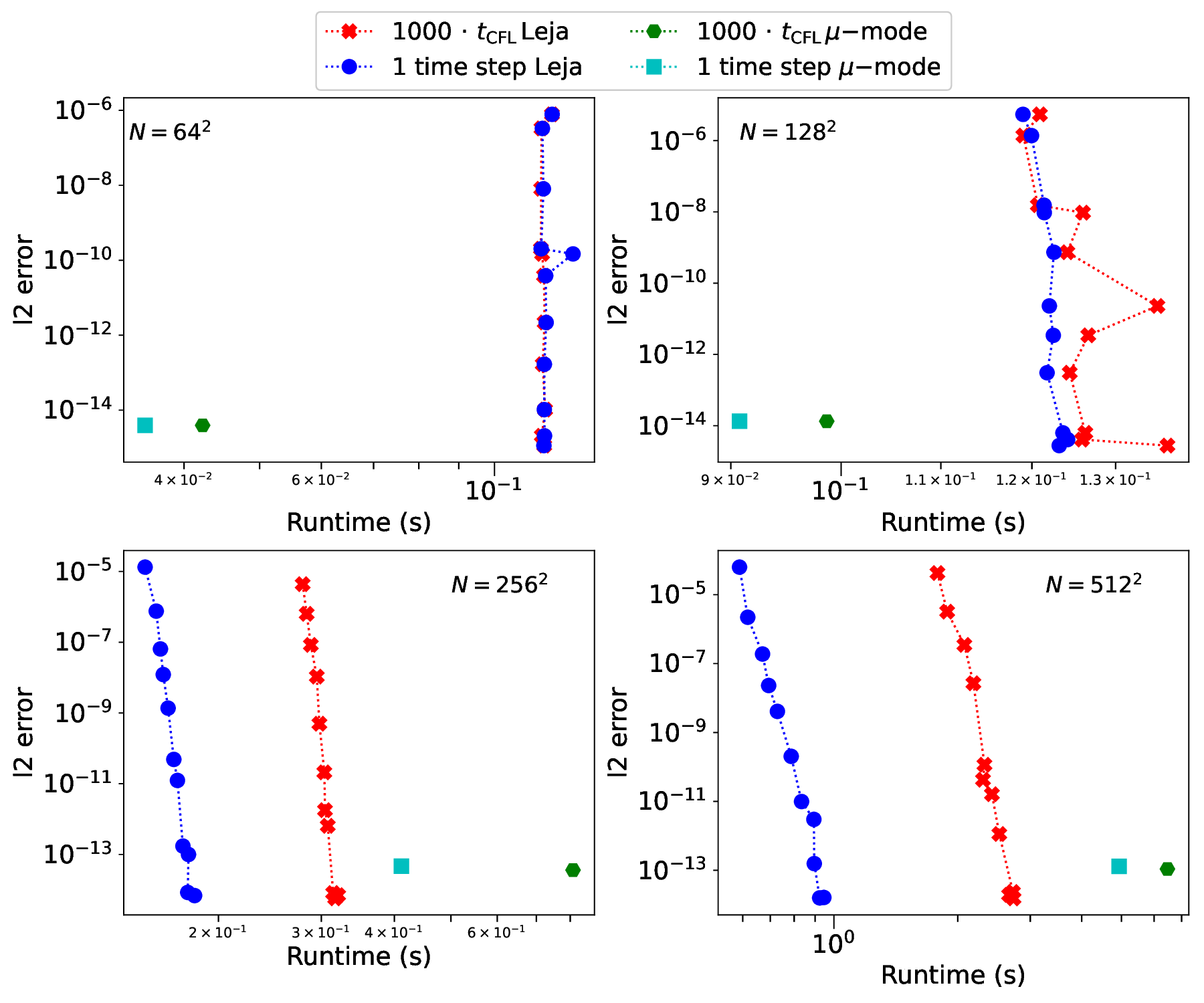}
    \caption{Comparison of the performance, i.e. l2 norm of the error incurred versus the computational runtime (in seconds), of the $\mu$-mode integrator with that of the Leja interpolation scheme for Case II at $T = 0.02$. The blue and red markers correspond to the error incurred for user-defined tolerances ranging from $10^{-2}$ to $10^{-12}$ (for Leja) in steps of an order of magnitude.}
    \label{fig:cost_mu_g_0.02}
\end{figure}

\begin{figure}[t!]
    \centering
    \includegraphics[width = \textwidth]{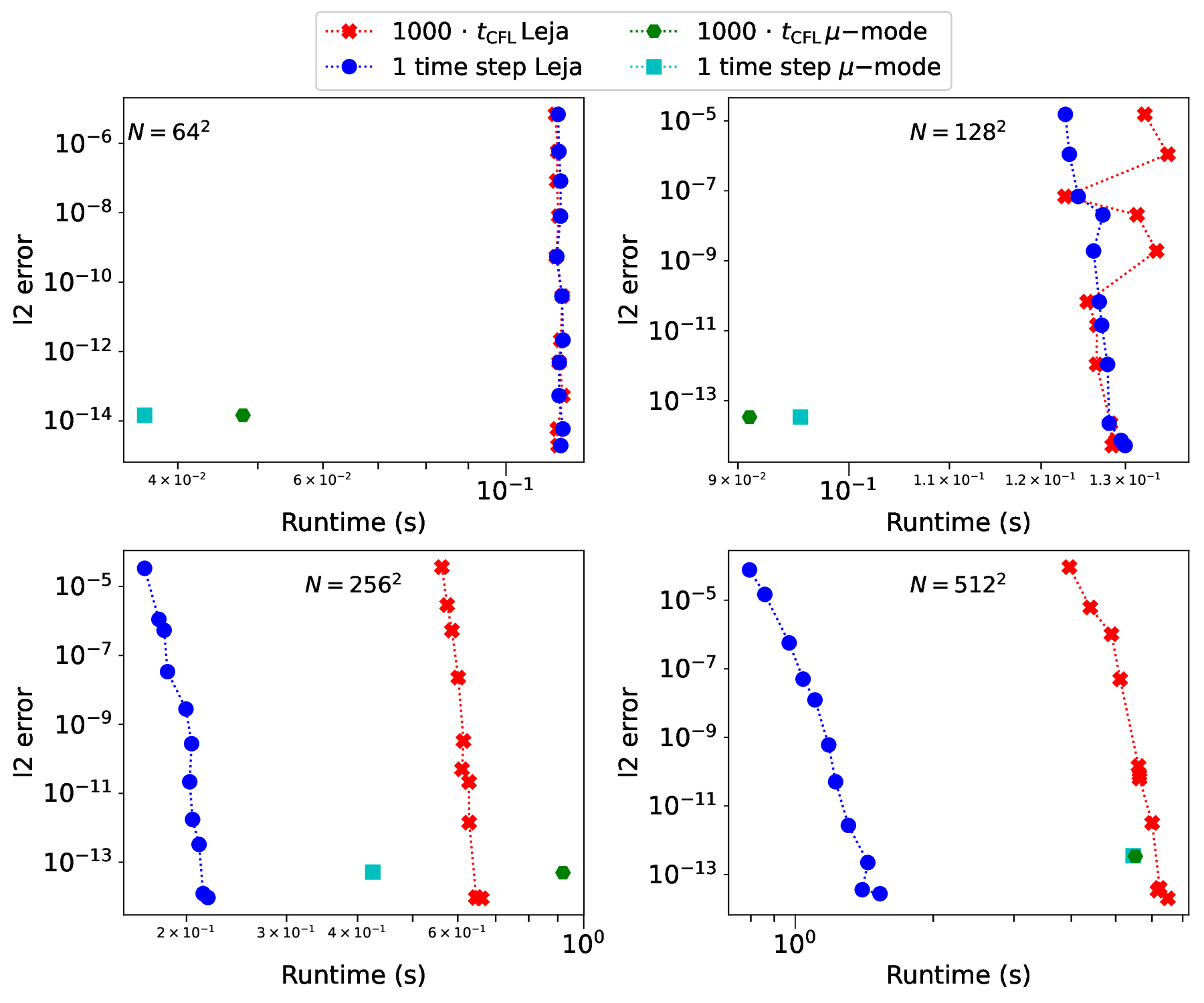}
    \caption{Same as Fig. \ref{fig:cost_mu_g_0.02}, but at $T = 0.05$.}
    \label{fig:cost_mu_g_0.05}
\end{figure}

\begin{figure}[t!]
    \centering
    \includegraphics[width = \textwidth]{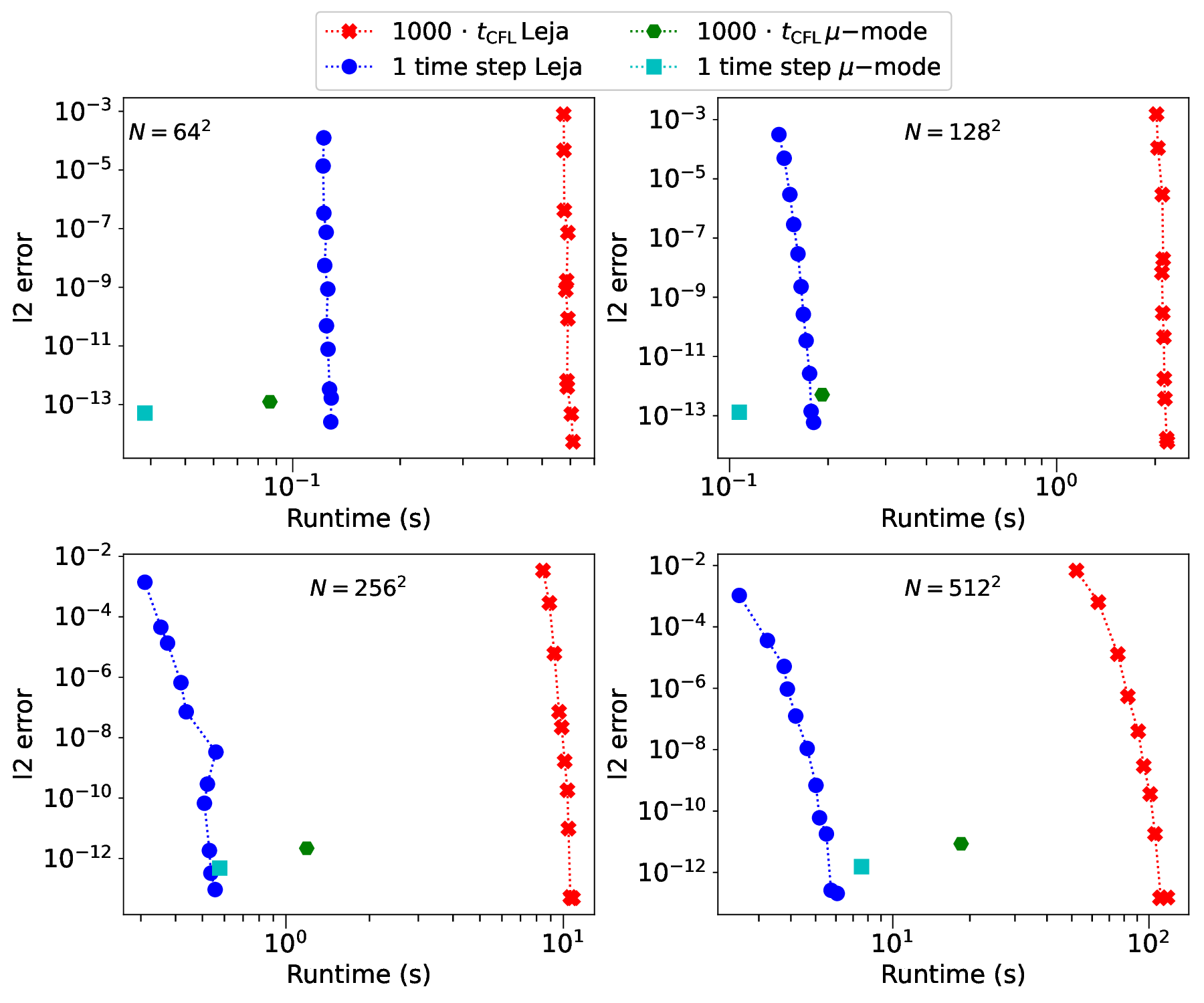}
    \caption{Same as Fig. \ref{fig:cost_mu_g_0.02}, but at $T = 1.0$ (steady-state solution).}
    \label{fig:cost_mu_g_1.0}
\end{figure}

In the test case of the diffusion of a Gaussian pulse along the X-direction (Case II), the mixed derivatives terms vanish. This allows us to directly apply the $\mu$-mode integrator. For constant step size implementation, one needs to compute a total of four matrix exponentials: one each for the two-dimensions at the start of the simulations (for a given value of $\Delta t$) and one each for each dimension at the final time step size (as the step size at the final time step is likely to differ from the initial step size). Similar to the Leja method, we consider two cases  (i) $\Delta t = 1000 \cdot \Delta t_\mathrm{CFL}$, and (ii) $\Delta t = T_f$. If we have multiple time steps, the computational cost incurred by the $\mu$-mode integrator is mainly dictated by the matrix-matrix products at every time step (Eq. \eqref{eq:mu_solution}). If, however, we have only one time step, the net cost incurred is given by the computation of four matrix exponentials using Pad\'e approximation and one matrix-matrix product. The cost incurred by Pad\'e approximation may depend on the step size under consideration. In the test problem considered in this work (Case II), we find that we can choose the step size to be large enough so that it is equal to the the simulation time without compromising the accuracy of the solution (Figs. \ref{fig:cost_mu_g_0.02}, \ref{fig:cost_mu_g_0.05}, and \ref{fig:cost_mu_g_1.0}). As the $\mu$-mode integrator does not require any matrix-vector products, we compare the computational cost in terms of the wall-clock time. These simulations have been conducted on an AMD Ryzen 5 1600 Six-Core Processor (64 bit). All simulations have been performed on a single thread. For $T_f = 0.02$ and $0.05$ and low resolution ($N = 64^2, 128^2$), choosing $\Delta t = 1000 \cdot t_\mathrm{CFL}$ is equivalent to choosing a single time step. In these cases, the $\mu$-mode integrator proves to be much more efficient than the Leja scheme. It has to be noted that in these cases, computing the coefficients of the polynomial using the divided differences algorithm, for the Leja scheme, has a substantial contribution to the overall computational expenses. This may be optimised by choosing to compute the divided differences for a fewer number of Leja points, say 1000 -- 2000 instead of the maximum available 10000. However, it is not straightforward to determine the number of Leja points needed for convergence prior to conducting the simulations. Hence, we compute the coefficients for all 10000 Leja points. With the increase in the number of grid points, we find that the Leja scheme, with $\Delta t = T_f$, tends to be slightly faster than the $\mu$-mode integrator.

Finally, we compare the performances of the ARK2, ARK4, and ETDRK2 integrators, with two different values of $\Delta t$, where we penalise the anisotropic diffusion equation by a Laplacian. We choose three cases, $T_f = 0.2, 4.0$ for the periodic band, and $T_f = 1.0$ for the Gaussian pulse, to investigate the performance of the penalisation approach as well as the ARK methods. This penalisation approach was tested out for the anisotropic diffusion problem in Ref. \cite{Crouseilles15} that gave a qualitative analysis of the performance of the two ARK integrators (amongst others). Our preliminary tests showed that the first-order exponential Euler integrator results in a hysteresis of diffusion. This is due to the artificially included parameter $\lambda$ that introduces additional diffusion (along both X-- and Y-- directions), which is then `extracted out' with an explicit integrator. A first-order integrator is unable to perform this extraction efficiently, and this results in the onset of a `delay' in the solution. This is in agreement with the results obtained in Ref. \cite{Crouseilles15}, and owing to this time lag, we choose to omit the first-order integrator from this work. Similar to the Crank--Nicolson solver, we use the GMRES iterative solver to compute the different stages in ARK2 and ARK4. The computational cost for these two integrators is dictated by the number of matrix-vector products. In ETDRK2, the non-stiff term is evaluated using the Leja interpolation scheme which constitutes the most expensive part of the computation. The constant coefficient linear term is evaluated using the $\mu$-mode integrator, and as seen from Figs. \ref{fig:cost_mu_g_0.02}, \ref{fig:cost_mu_g_0.05}, and \ref{fig:cost_mu_g_1.0}, the cost incurred by this integrator is roughly similar to the case of directly computing the matrix exponential using the Leja approach. This cost is negligible, at least more than an order of magnitude and up to almost three orders of magnitude smaller compared to the cost of the computing the remainder, i.e. multiple $\varphi_l$ functions per time step, using the Leja scheme. This is why we only account for the matrix-vector products needed for Leja interpolation and not for the matrix-matrix products in the $\mu$-mode integrator in determining the cost incurred by ETDRK2.

\begin{figure}[t!]
    \centering
    \includegraphics[width = \textwidth]{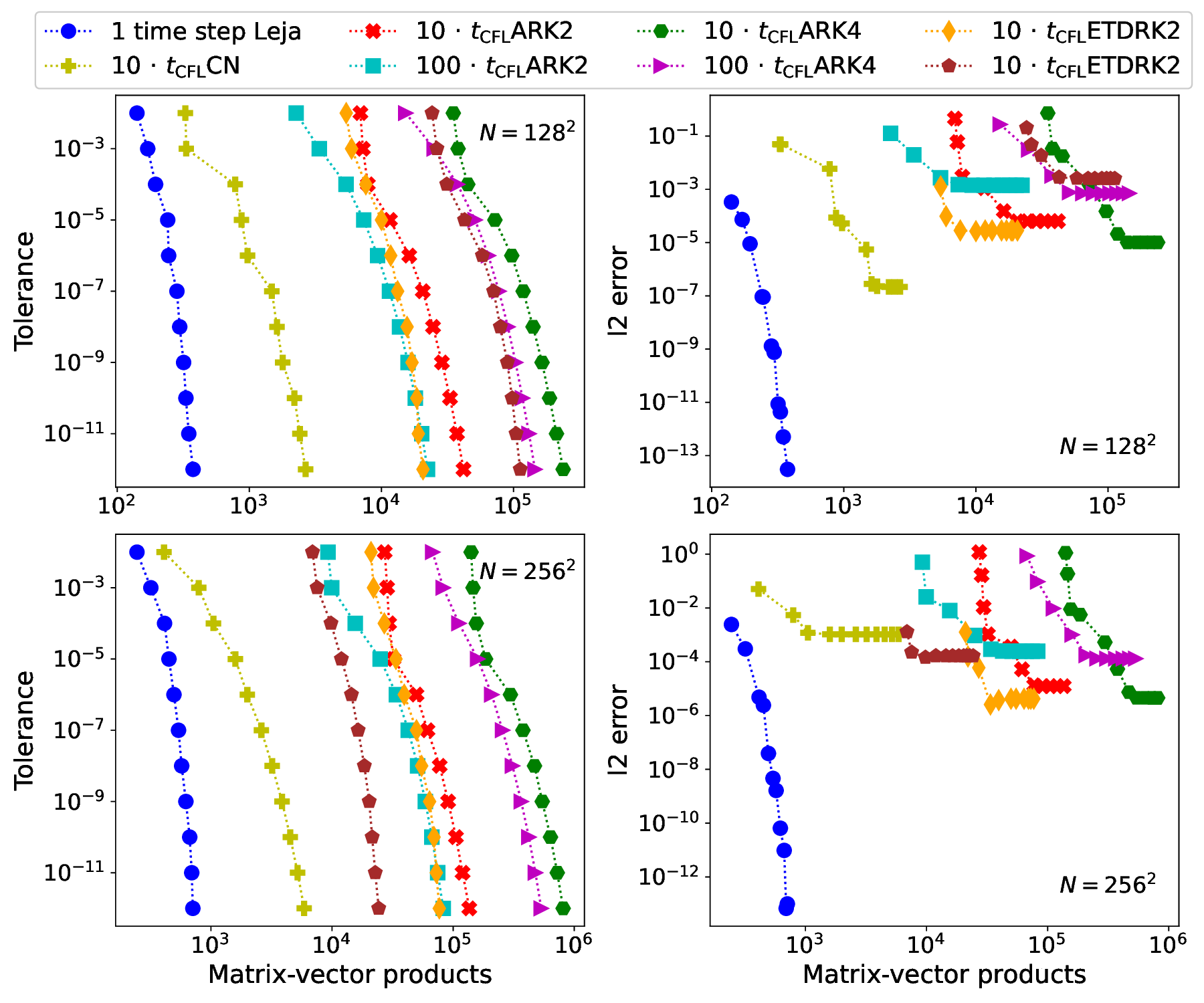}
    \caption{Performance comparison of ARK2, ARK4, and ETDRK2, for Case I at $T = 0.2$. The Leja scheme ($\Delta t = T_f$) and Crank--Nicolson ($\Delta t = 10 \cdot t_\mathrm{CFL}$) are added for reference. The left panel shows the user-defined tolerance whereas the right panel shows the l2 norm of the error incurred, as a function of the computational time for different spatial resolutions.}
    \label{fig:cost_ark_b_0.2}
\end{figure}

\begin{figure}[t!]
    \centering
    \includegraphics[width = \textwidth]{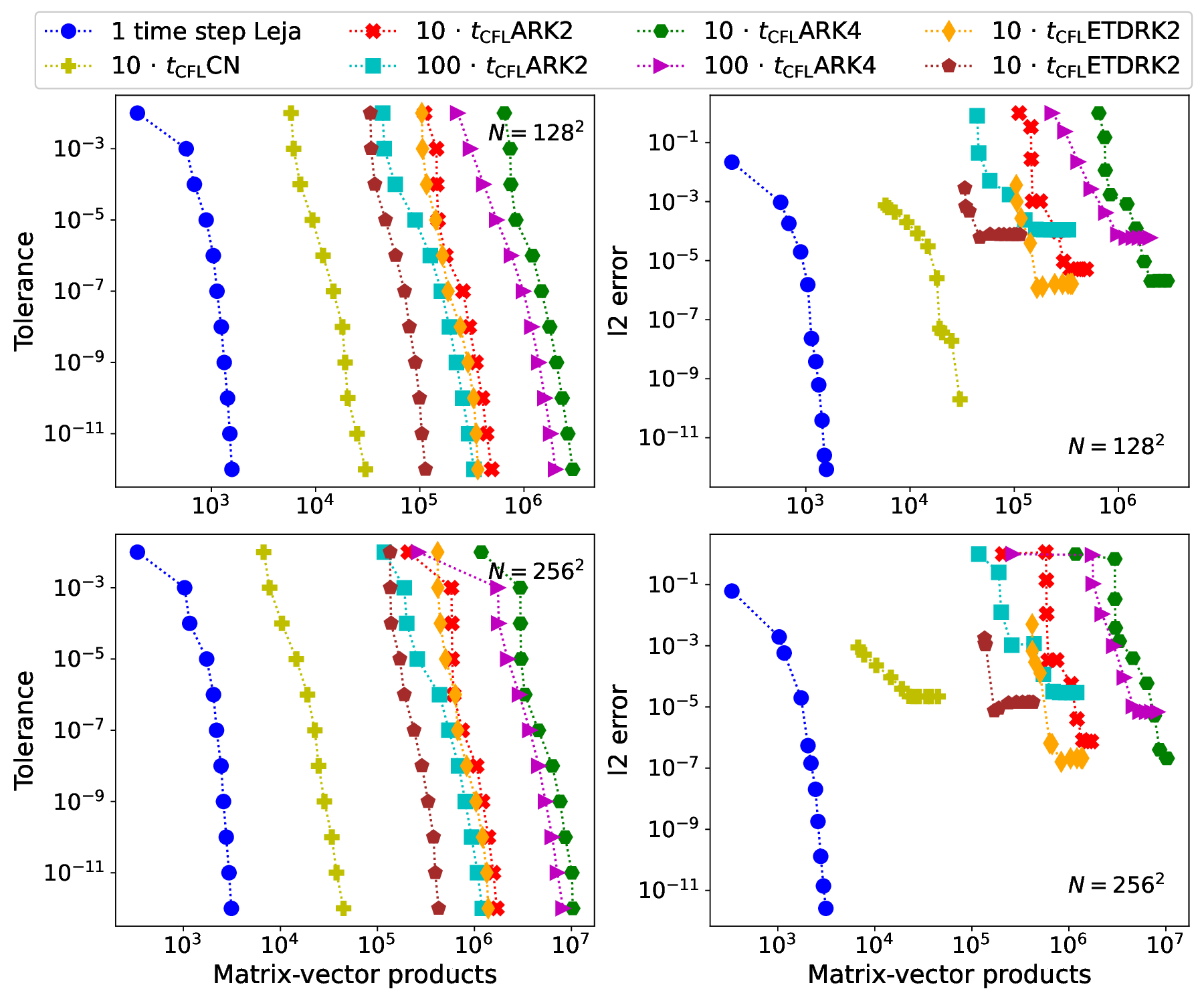}
    \caption{Performance comparison of ARK2, ARK4, and ETDRK2, for Case I at $T = 4.0$ (steady state solution). The Leja scheme ($\Delta t = T_f$) and Crank--Nicolson ($\Delta t = 10 \cdot t_\mathrm{CFL}$) are added for reference. The left panel shows the user-defined tolerance whereas the right panel shows the l2 norm of the error incurred, as a function of the computational time for different spatial resolutions.}
    \label{fig:cost_ark_b_4.0}
\end{figure}

\begin{figure}[t!]
    \centering
    \includegraphics[width = \textwidth]{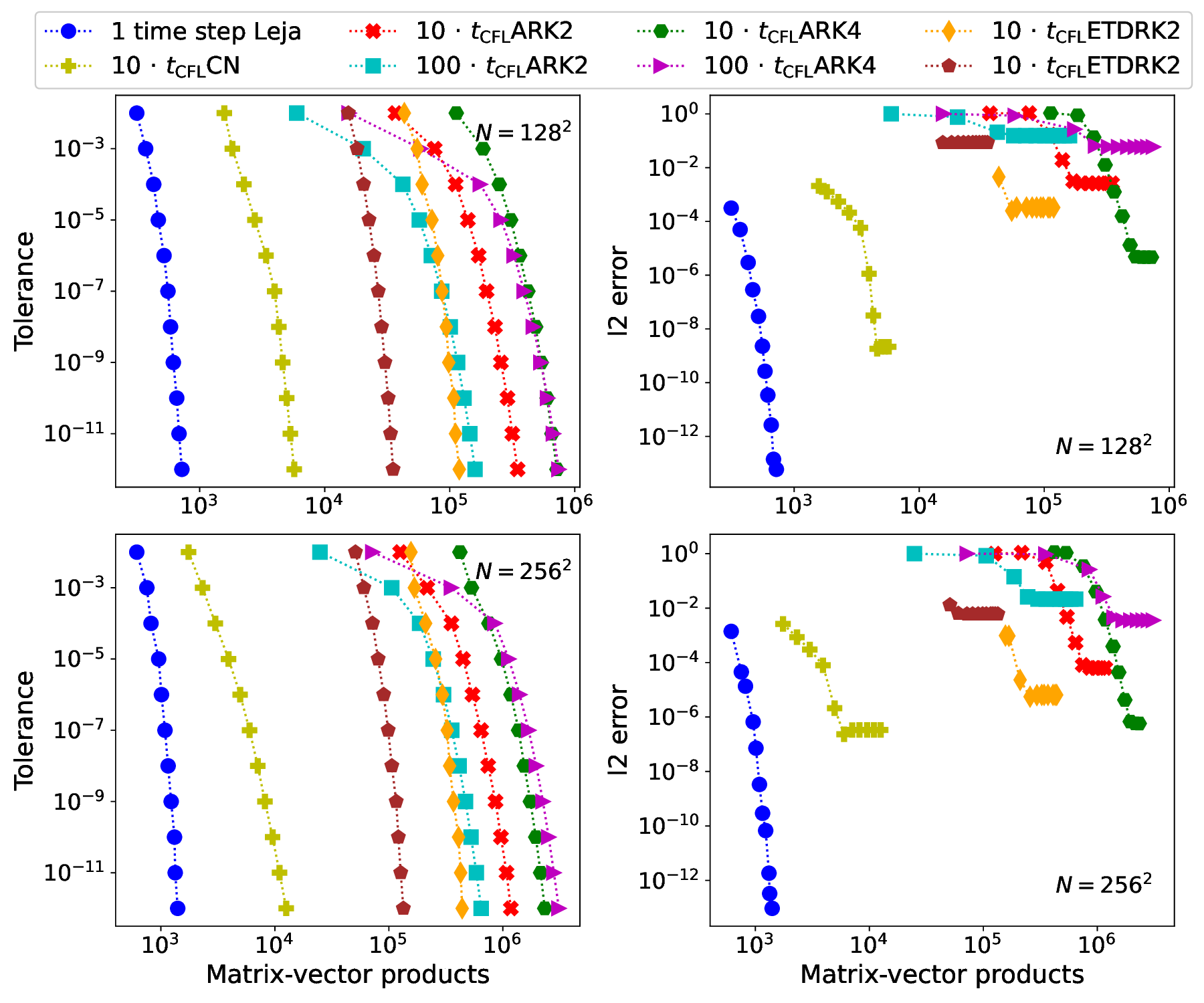}
    \caption{Performance comparison of ARK2, ARK4, and ETDRK2, for Case II at $T = 1.0$ (steady state solution). The Leja scheme ($\Delta t = T_f$) and Crank--Nicolson ($\Delta t = 10 \cdot t_\mathrm{CFL}$) are added for reference. The left panel shows the user-defined tolerance whereas the right panel shows the l2 norm of the error incurred, as a function of the computational time for different spatial resolutions.}
    \label{fig:cost_ark_g_1.0}
\end{figure}

Figs. \ref{fig:cost_ark_b_0.2}, \ref{fig:cost_ark_b_4.0}, and \ref{fig:cost_ark_g_1.0} illustrate that the penalisation approach is extremely expensive compared to that of the Leja method (between two to four orders of magnitude) whilst being unable to generate accurate results. We observe that choosing smaller step sizes can lead to better results for ARK2 and ARK4 but one has to pay the price of an increased computational cost. This is true for all values of user-specified tolerance. For similar step sizes, the fourth-order integrator yields more accurate results than the second-order one. For a given step size, ETDRK2 yields more accurate results than the ARK methods whilst being cheaper by up to an order of magnitude. Higher-order ETDRK methods may be able to generate even more precise results but they require several computations of the $\varphi_l$ functions.

The choice of the penalisation parameter, $\lambda$, plays a significant role. On one hand, large values of $\lambda$ render the system extremely stiff and the cost of conducting the simulations increases dramatically. On the other hand, small values of $\lambda$ may lead to unstable solutions, especially in the case of integrators with multiple internal stages. We find that using small values of $\lambda  \, (= 2 |\alpha|)$, with ARK4, lead to violent oscillations in the distribution function. However, similar values of $\lambda$ work reasonably well for ARK2 and ETDRK2. 

Furthermore, we would like note that the extraordinary expense of the ARK and ETDRK2 methods using the penalisation approach (as compared to Leja or Crank--Nicolson), is due to the fact it is more expensive to invert the Laplacian operator than the \textbf{anisotropic} diffusion operator. This is because the anisotropy in diffusion actually scales down the magnitude of the elements in the matrix by the diffusion coefficients (as they are less than 1). E.g., in the case of the diffusion of a Gaussian in a periodic band, the diffusion coefficient along the Y-direction is $0.25$, which means that the components of the anisotropic diffusion matrix, along the Y-direction, are scaled down to 0.25. In the case of the Gaussian pulse, $D_{yy} = 0$, and the anisotropic diffusion matrix becomes even more sparse. In the Laplacian matrix, however, there is no scaling down of the elements of the matrix, which is why, it is more expensive to invert this operator.

For a linear differential equation, as the method of the direct computation of the matrix exponential using Leja interpolation is not bounded by stability or accuracy constraints, one could expect this method to always surpass the others. The penalisation approach is a less accurate and less stable scheme, and therefore, one has to use smaller time step sizes, consequently resulting in an enormous number of time steps (and therefore, iterations). However, in situations where we have are forced to choose smaller time steps for the Leja method/exponential integrators, e.g. equations involving time-dependent sources or some nonlinear terms, the issue of less accuracy is mitigated as the time step size is limited by the physics of the problem (i.e. how fast the source term changes, nonlinearities, etc). This is where we would expect the penalisation approach, with a different penalisation operator, to show substantial computational improvements. This will be a subject of our future work.

\section{Conclusions}
\label{sec:conclude}

Numerically solving the anisotropic diffusion equation forms a crucial component in understanding the physics of CR diffusion as well as properties of the Galactic magnetic field. The stiff nature of the linear diffusion equation can be combated by directly computing the exponential of the matrix, instead of using implicit (or explicit) methods, which allows for one to take extremely large step sizes without comprising the accuracy of the solution. In certain cases, this may even allow us to take only a single time step, thereby expediting the simulations by a colossal margin.

Using a number of test models for anisotropic diffusion, we show that the iterative method of polynomial interpolation at Leja points and the direct approach of the $\mu$-mode integrator, that makes use of the Kronecker form of the underlying matrices, significantly outperform any implicit or semi-implicit integrator. We also test the performance of a penalisation approach with semi-implicit and exponential integrators. We find that this approach is computationally intensive as we artificially add extra diffusion and try to treat the remainder term using some explicit methods. 

Even though all examples considered in this study are two-dimensional, the extension to three dimensions is rather straightforward for the Leja interpolation method, exponential integrators, and Crank--Nicolson. However, the situation for the direct solver, i.e. the $\mu$-mode integrator, becomes a bit more complicated. The cost of computing the matrix exponentials scales as $O(\mathrm{N}^3)$, where `$\mathrm{N}$' is the number of grid points in each direction. The cost of the matrix-matrix products scales as $O(\mathrm{N} \cdot \mathrm{N}^2)$, which in 2D, is roughly the same as computing the exponentials of the matrices. In 3D, however, the cost of computing the exponentials would still scale as $O(\mathrm{N}^3)$, whereas the matrix-matrix products scale as $O(\mathrm{N}^4)$, and this is expected to dictate the overall computational cost.

In our companion work \cite{Kis22}, we compare the performance of the alternating-direction implicit and Crank--Nicolson methods (traditionally used in \textsc{GalProp}) against our proposed method of computing the matrix exponential using the Leja method within the framework of \picard. This has been considered for toy models with pure isotropic CR diffusion, diffusion plus advection, and cases including continuous energy losses in the presence of time-dependent CR sources. We show that the Leja method outperforms both these implicit methods in computational effort as well as accuracy by a substantial margin. The Leja method is well suited for treating realistic time-dependent CR transport problems, whilst accounting for anisotropic diffusion. Our future work, in this direction, is to test the efficacy of the Leja scheme in anisotropic diffusion plus advection and (qualitatively) more realistic diffusion tensors against higher order implicit and semi-implicit methods. It will be interesting to see if the Leja method would be able to achieve such considerable speed-ups in cases where the imaginary eigenvalues of the underlying matrices are non-negligible. Furthermore, we will consider cases of time-varying diffusion coefficients that would emulate time-varying magnetic fields in the form of sinusoidal variations or exponential decay as well as moving magnetic fields. The applicability of the $\mu$-mode integrator in such convoluted scenarios remains to be seen. We will also investigate the possibility of extending the $\mu$-mode integrator to handle matrices that are not expressible in the form of the Kronecker sum, i.e., $A_x \otimes A_y$. Such an approach will allow us to treat the remainder term, in the penalisation method, exponentially.


\section*{Acknowledgements}
This work has been supported, in part, by the Austrian Science Fund (FWF) project id: P32143-N32. PJD would like to thank Fabio Cassini for the useful discussions related to the $\mu$-mode integrator.


\appendix






\bibliographystyle{elsarticle-harv} 
\bibliography{ref}





\end{document}